\pdfoutput=1
\documentclass[11pt]{article}

\PassOptionsToPackage{hidelinks}{hyperref}
\usepackage[final]{acl}

\usepackage{times}
\usepackage{latexsym}

\usepackage[T1]{fontenc}
\usepackage[utf8]{inputenc}

\usepackage{microtype}

\usepackage{inconsolata}

\usepackage{graphicx}
\usepackage{float}

\usepackage{subfigure}   % legacy API you’re using; fine if you keep it
\usepackage{booktabs}
\usepackage{multirow}
\usepackage{tabularx}
\usepackage{tabulary}
\usepackage{threeparttable}
\usepackage{tablefootnote}
\usepackage{array}
\renewcommand\arraystretch{1.3}
\newcolumntype{C}[1]{>{\PreserveBackslash\centering}p{#1}}
\newcolumntype{R}[1]{>{\PreserveBackslash\raggedleft}p{#1}}
\newcolumntype{L}[1]{>{\PreserveBackslash\raggedright}p{#1}}

\usepackage{tikz}
\usepackage[edges]{forest}

\definecolor{hidden-draw}{RGB}{20,68,106}
\definecolor{hidden-pink}{RGB}{255,245,247}
\definecolor{mattered}{RGB}{214,26,60}
\definecolor{mattegreen}{HTML}{369F39}

\usepackage{enumitem}
\usepackage{nicefrac}
\usepackage{xurl}        % better URL line breaks
\usepackage{wrapfig}
\usepackage{amsmath,amsfonts}
\usepackage{dsfont}
\usepackage{mathtools}
\usepackage{bm}
\usepackage{algorithm}
\usepackage{algpseudocode}
\usepackage{fontawesome5}
\usepackage{natbib}
\usepackage{xcolor}
\usepackage{tikz}
\usepackage[most]{tcolorbox}
\usepackage{hyperref}
\title{A Comprehensive Survey on the Trustworthiness of Large Language Models in Healthcare}

\author{Manar Aljohani \\
  Virginia Tech / Address line 1 \\
  Affiliation / Address line 2 \\
  Affiliation / Address line 3 \\
  \texttt{email@domain} \\\And
  Xuan Wang \\
  Virginia Tech / Address line 1 \\
  Affiliation / Address line 2 \\
  Affiliation / Address line 3 \\
  \texttt{email@domain} \\}

\author{
  Manar Aljohani \textsuperscript{$\spadesuit$},
  Jun Hou \textsuperscript{$\spadesuit$},
  Sindhura Kommu \textsuperscript{$\spadesuit$},
  Xuan Wang \textsuperscript{$\spadesuit$}\\
  \textsuperscript{$\spadesuit$} Department of Computer Science, Virginia Tech, Blacksburg, VA, USA\\
  \texttt{\{manara, junh, sindhura, xuanw\}@vt.edu}
}
\begin{document}
\maketitle
\begin{abstract}
The application of large language models (LLMs) in healthcare holds significant promise for enhancing clinical decision-making, medical research, and patient care. However, their integration into real-world clinical settings raises critical concerns around trustworthiness, particularly around dimensions of truthfulness, privacy, safety, robustness, fairness, and explainability. These dimensions are essential for ensuring that LLMs generate reliable, unbiased, and ethically sound outputs. While researchers have recently begun developing benchmarks and evaluation frameworks to assess LLM trustworthiness, the \textbf{trustworthiness of LLMs in healthcare} remains underexplored, lacking a systematic review that provides a comprehensive understanding and future insights. This \textbf{survey} addresses that gap by providing a comprehensive review of current methodologies and solutions aimed at mitigating risks across key trust dimensions. We analyze how each dimension affects the reliability and ethical deployment of healthcare LLMs, synthesize ongoing research efforts, and identify critical gaps in existing approaches. We also identify emerging challenges posed by evolving paradigms, such as multi-agent collaboration, multi-modal reasoning, and the development of small open-source medical models. Our goal is to guide future research toward more trustworthy, transparent, and clinically viable LLMs.
\end{abstract}

\begin{figure*}[t]
  \centering
  \subfigure[Temporal Trends ]{
    \includegraphics[width=0.31\textwidth]{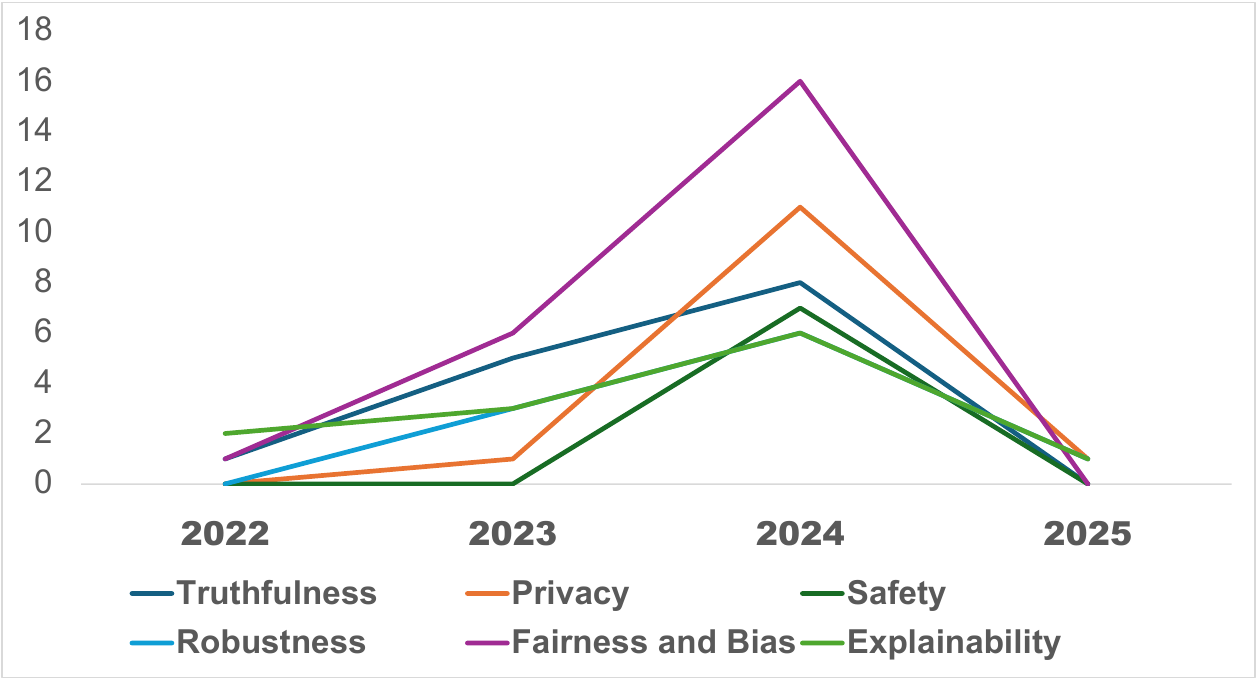}
    \label{fig:papersYear}
  }
  \hfill
  \subfigure[Distribution of Datasets]{
    \includegraphics[width=0.32\textwidth]{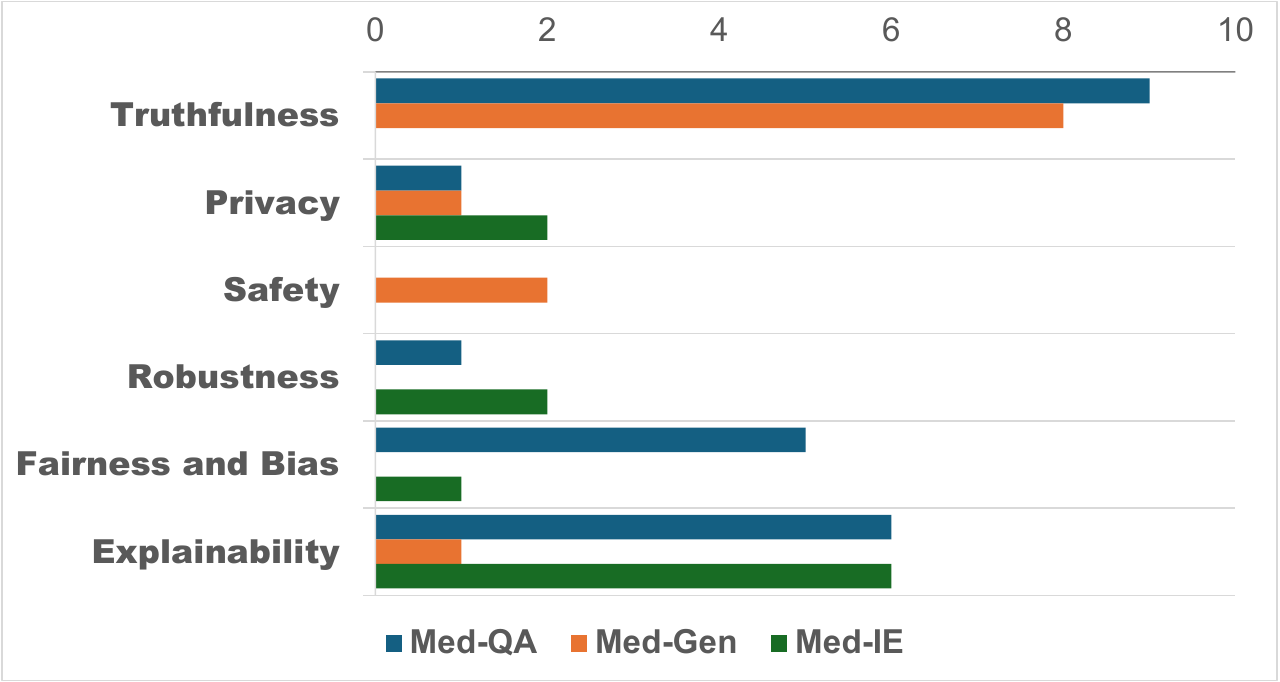}
    \label{fig:datasetsTask}
  }
  \hfill
  \subfigure[Distribution of Models]{
    \includegraphics[width=0.31\textwidth]{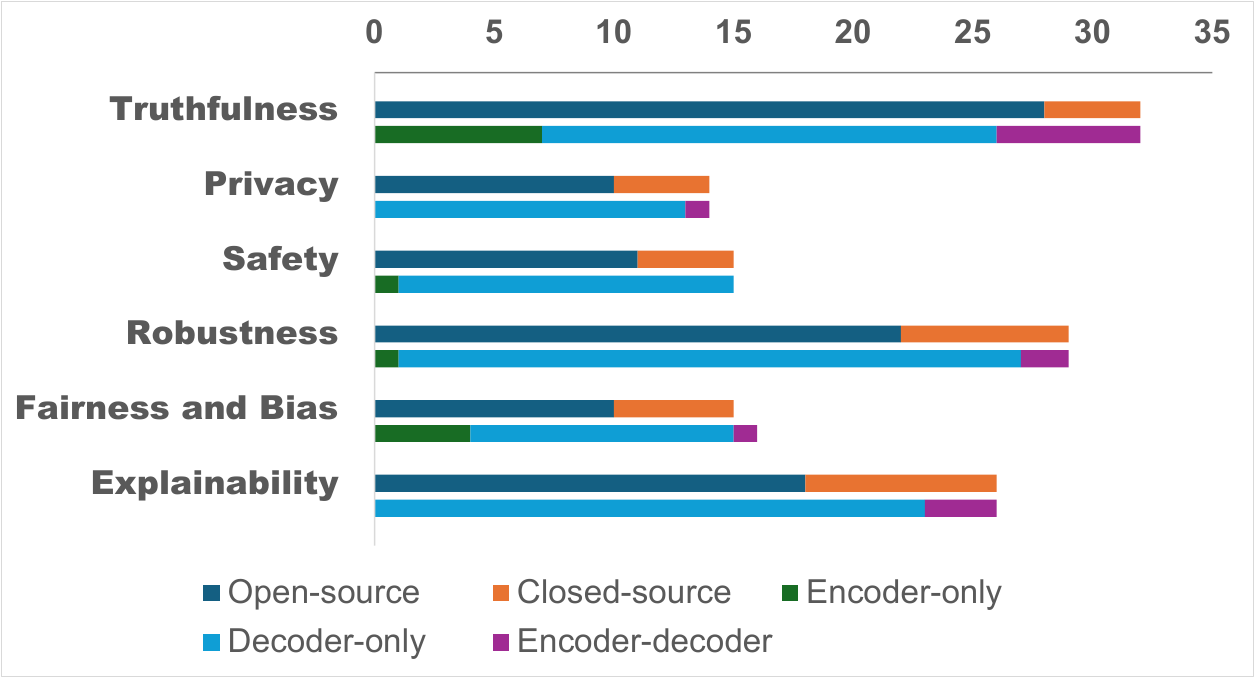}
    \label{fig:modelsDimensions}
  }
  \caption{Overview of research trends, dataset usage, and model types across key trustworthiness dimensions in healthcare LLM studies: (a) Temporal Trends in Trustworthiness Dimensions Addressed in Medical LLM Studies (2022–2025); (b) Distribution of Dataset Types Across Trustworthiness Dimensions in Healthcare LLM Studies; (c) Distribution of Model Types Across Trustworthiness Dimensions in Healthcare LLM Studies.}
 % \label{fig:combined_summary}
\end{figure*}

\section{Introduction}
The application of LLMs in healthcare is advancing rapidly, with the potential to transform clinical decision-making, medical research, and patient care. However, incorporating them into healthcare systems poses several key challenges that need to be addressed to ensure their reliable and ethical use. As highlighted in \citet{bi2024ai}, a major concern is the trustworthiness of AI-enhanced biomedical insights. This encompasses improving model explainability and interpretability, enhancing robustness against adversarial attacks, mitigating biases across diverse populations, and ensuring strong data privacy protections. Key concerns include truthfulness, privacy, safety, robustness, fairness, and explainability, each of which plays a vital role in the reliability and trustworthiness of AI-driven healthcare solutions. 

\emph{Truthfulness}, defined as "the accurate representation of information, facts, and results by an AI system" \cite{huang2024position}, is critical in healthcare, as inaccuracies can lead to misdiagnoses or inappropriate treatment recommendations. Ensuring that generated information is both accurate and aligned with verified medical knowledge is essential. Additionally, \emph{privacy} concerns arise from the risk of exposing sensitive patient data during model training and usage, potentially leading to breaches or violations of regulations such as HIPAA (Health Insurance Portability and Accountability Act) and GDPR (General Data Protection Regulation). Ensuring patient confidentiality while leveraging LLMs for diagnostics and treatment recommendations is a critical challenge. \emph{Safety}, defined as “ensuring that LLMs do not answer questions that can harm patients or healthcare providers in healthcare settings” \cite{han2024medsafetybench}, further underscores the necessity of implementing stringent safeguards to mitigate harm. \emph{Robustness} refers to an LLM’s ability to consistently generate accurate, reliable, and unbiased outputs across diverse clinical scenarios while minimizing errors, hallucinations, and biases. It also encompasses the model’s resilience against adversarial attacks, ensuring that external manipulations do not compromise its integrity. A truly robust LLM in healthcare must demonstrate stability, reliability, and fairness, even when faced with noisy, ambiguous, or adversarial inputs. Similarly, \emph{fairness and bias} must be addressed to prevent discriminatory patterns in model outputs, which could lead to unequal treatment recommendations and exacerbate healthcare disparities. Furthermore, the \emph{explainability} of LLMs, which ensures that model outputs are interpretable and transparent, plays a vital role in fostering trust and allowing informed decision-making by healthcare professionals. Lack of transparency in model reasoning complicates clinical adoption and raises accountability concerns. 

Clinical deployments of LLMs expose trust gaps across dimensions. Med-PaLM and Med-PaLM 2 show truthfulness and safety issues, with hallucinated that could misguide care \cite{singhal2023large}. Integrating LLMs with EHRs in cloud settings risks HIPAA/GDPR violations, prompting on-prem deployment and stronger de-identification \cite{jonnagaddala2025privacy}. Robustness remains problematic; frameworks like MEDIC and CREOLA assess hallucination severity and clinical safety \cite{kanithi2024medic, asgari2025framework}. Fairness issues persist, with studies showing that LLMs can perpetuate racial biases in medical recommendations \cite{pfohl2024toolbox}. Finally, explainability challenges were evident in AMIE—a conversational diagnostic agent evaluated in OSCE-style clinical exams—demonstrated strong diagnostic reasoning but lacked transparency compared to human doctors \cite{tu2025towards}.

Tackling these challenges is essential for the trustworthy and ethical implementation of LLMs in healthcare. Recently, researchers have begun developing benchmarks and evaluation frameworks to systematically assess the trustworthiness of LLMs \cite{huang2024position}. The \textbf{trustworthiness of LLMs in healthcare} is gaining increasing attention due to its significant social impact. However, there is currently no systematic review that provides a comprehensive understanding and future insights into this area. To bridge this gap, we present a comprehensive \textbf{survey} that explores these trust-related dimensions in detail, reviewing existing benchmarks and methodologies aimed at improving the trustworthiness of LLMs in healthcare.

\section{Datasets, Models, and Tasks}
\subsection{Inclusion \& Exclusion Criteria} 
We initiated our survey with a comprehensive literature search targeting studies on the trustworthiness of LLMs in healthcare. Our search strategy employed diverse keyword combinations and was directed toward top-tier conferences and journals, prioritizing publications from 2022 onward. Detailed inclusion and exclusion criteria are provided in Appendix~\ref{app:InclusionExclusionCriteriaDetails}. Fig~\ref{fig:papersYear} illustrates how the number of papers addressing each key trustworthiness dimension in healthcare LLMs has changed over time from 2022 to 2025. From Figure \ref{fig:papersYear}, interest in trustworthiness dimensions peaked in 2024, particularly for Fairness and Bias (16 papers) and Privacy (11 papers), reflecting a strong recent push toward ethical and secure AI in healthcare. Truthfulness and Explainability maintained steady growth through 2023 and 2024. These trends suggest a rising concern with fairness and privacy in recent years, possibly driven by real-world deployment risks and regulatory pressure.

% We then summarize all the datasets, models, and tasks relevant to research on trust in LLMs for healthcare, providing a comprehensive overview of their applications and contributions to this domain. 

\tikzstyle{my-box}=[
    rectangle,
    draw=hidden-draw,
    rounded corners,
    text opacity=1,
    minimum height=1.5em,
    minimum width=5em,
    inner sep=2pt,
    align=center,
    fill opacity=.5,
    line width=0.8pt,
]
\tikzstyle{leaf}=[my-box, minimum height=1.5em,
    fill=hidden-pink!80, text=black, align=left,font=\normalsize,
    inner xsep=2pt,
    inner ysep=4pt,
    line width=0.8pt,
]

% Don't leave line space between branches 
\begin{figure*}[t!]
    \centering
    \resizebox{\textwidth}{!}{
        \begin{forest}
            forked edges,
            for tree={
                grow=east,
                reversed=true,
                anchor=base west,
                parent anchor=east,
                child anchor=west,
                base=center,
                font=\large,
                rectangle,
                draw=hidden-draw,
                rounded corners,
                align=left,
                text centered,
                minimum width=4em,
                edge+={darkgray, line width=1pt},
                s sep=3pt,
                inner xsep=2pt,
                inner ysep=3pt,
                line width=0.8pt,
                ver/.style={rotate=90, child anchor=north, parent anchor=south, anchor=center},
            },
            where level=1{text width=14em,font=\normalsize,}{},
            where level=2{text width=18em,font=\normalsize,}{},
            where level=3{text width=18em,font=\normalsize,}{},
            [
                \textbf{Trustworthiness of LLMs in Healthcare}, ver
                                 [
                                \textbf{Truthfulness}, fill=blue!10
                            [
                                \textbf{Benchmarks}, fill=blue!10
                            [
                             {Med-HALT~\cite{pal-etal-2023-med}, PubHealthTab~\cite{akhtar-etal-2022-pubhealthtab}, HEALTHVER~\cite{sarrouti-etal-2021-evidence-based}}, leaf, text width=66em
                            ]
                                ]
                                [
                                \textbf{Mitigation Methods}, fill=blue!10
                            [
                         {Self Reflection~\cite{ji-etal-2023-towards}{,}  MEDAL~\cite{li-etal-2024-better}{,} Faithful Reasoning~\cite{tan2024faithful}{,} HEALTHVER~\cite{sarrouti-etal-2021-evidence-based}{,}\\ CRITIC~\cite{gou2024critic}{,} SEND~\cite{mohammadzadeh2024hallucination}}, leaf, text width=66em
                            ]
                                ]
                             [
                                \textbf{Evaluation and Detection Methods}, fill=blue!10
                            [
                        {Med-HALT~\cite{pal-etal-2023-med}{,} Med-HVL~\cite{yan2024med}{,}
                        Semantic Entropy~\cite{farquhar2024detecting}{,}\\ SEPs~\cite{han2024semantic}{,} Faithful Reasoning~\cite{tan2024faithful}{,} PubHealthTab~\cite{akhtar-etal-2022-pubhealthtab}{,} HEALTHVER~\cite{sarrouti-etal-2021-evidence-based}{,}\\ CRITIC~\cite{gou2024critic}{,} Cross-Examination~\cite{cohen2023lm}{,} MAD~\cite{smit2023we}}, leaf, text width=66em
                            ]
                                ]
                                ]
                        [
                            \textbf{Privacy}, fill=blue!10
                            [
                            \textbf{Benchmarks}, fill=blue!10
                            [ SecureSQL~\cite{song-etal-2024-securesql}, leaf, text width=66em
                            ]  
                        ]
                        [
                            \textbf{Enhance Methods}, fill=blue!10
                            [
                        {Federated Learning~\cite{zhao2024llm}{,}\\ 
                        Differential Privacy~\cite{singh2024whispered}{,} De-identification~\cite{liu2023deid}{,} Mitigating Memorization~\cite{yang2024memorization}{,} APNEAP~\cite{wu-etal-2024-mitigating-privacy}}, leaf, text width=66em
                            ] 
                        ]
                        [
                            \textbf{Evaluation Methods}, fill=blue!10
                            [
                        {SecureSQL~\cite{song-etal-2024-securesql}{,} Memorize Fine-tuning Data~\cite{yang2024memorization}{,} clinical Note De-identification~\cite{altalla2025evaluating}{,}\\ Memorization~\cite{yang2024memorization}{,} Textual Data Sanitization~\cite{xin2024a}}, leaf, text width=66em
                            ]
                        ]
                        ]
                        [
                            \textbf{Safety}, fill=blue!10
                            [
                            \textbf{Benchmarks}, fill=blue!10
                            [
                        {Med-harm~\cite{han2024towards}{,} Medsafetybench~\cite{han2024medsafetybench}}, leaf, text width=66em
                            ] 
                        ]
                        [
                            \textbf{Enhance Methods}, fill=blue!10
                            [ {UNIWIZ~\cite{das2024uniwiz}{,} Data-Poisoning Attack~\cite{han2024dataPoisoning}}, leaf, text width=66em
                            ]
                        ]
                        [
                            \textbf{Evaluation Methods}, fill=blue!10
                            [
                        {Med-harm~\cite{han2024towards}{,} Medsafetybench~\cite{han2024medsafetybench}{,} Misinformation Attacks~\cite{han2024medical}{,} MEDIC~\cite{kanithi2024mediccomprehensiveframeworkevaluating}{,}\\ GLiR Attack~\cite{leemann2024is}{,} Data-Poisoning Attack~\cite{han2024dataPoisoning}}, leaf, text width=66em
                            ]   
                        ]   
                        ]
                        [
                        \textbf{Robustness}, fill=blue!10
                        [
                        \textbf{Benchmarks}, fill=blue!10
                        [
                          {Detecting Anomalies~\cite{rahman2024incorporating}{,} RobustQA~\cite{han-etal-2023-robustqa}{,} RABBITS~\cite{gallifant-etal-2024-language}}, leaf, text width=66em
                        ]
                        ]
                        [
                        \textbf{Enhance Methods}, fill=blue!10
                        [
                         {LLM-TTA~\cite{o2024improving}{,} Detecting Anomalies~\cite{rahman2024incorporating}{,} Secure Your Model~\cite{tang-etal-2024-secure}{,} MEDSAGE~\cite{binici2025medsage}{,}\\ Out-of-Context Prompting~\cite{cotta2024out}}, leaf, text width=66em
                        ]
                        ]
                        [
                        \textbf{Evaluation Methods}, fill=blue!10
                        [
                         {Stumbling Blocks~\cite{wang2024stumbling}{,} Detecting Anomalies~\cite{rahman2024incorporating}{,}\\ Instruction Phrasings~\cite{ceballos-arroyo-etal-2024-open}{,} RobustQA~\cite{han-etal-2023-robustqa}{,} RABBITS~\cite{gallifant-etal-2024-language}}, leaf, text width=66em
                        ]
                        ]
                        ]
                        [
                        \textbf{Fairness and Bias}, fill=blue!10
                                 [
                        \textbf{Benchmarks}, fill=blue!10
                                 [ {BiasMedQA~\cite{schmidgall2024evaluation}{,} EquityMedQA~\cite{pfohl2024toolbox}{,} Superficial Fairness Alignment~\cite{wei2024actions}{,} FairMedFM~\cite{jin2024fairmedfm}}, leaf, text width=66em
                                 ]      
                        ]
                        [
                        \textbf{Mitigation Methods}, fill=blue!10
                                 [ {BiasMedQA~\cite{schmidgall2024evaluation}{,} Reinforcement Learning with Clinician Feedback~\cite{zack2024assessing}{,} Instruction Fine-tuning~\cite{singhal2023large}{,}\\ Hurtful Words~\cite{zhang2020hurtful}{,} Mitigate Cognitive Biases~\cite{ke2024mitigating}{,} CI4MRC~\cite{zhu2023causal}{,} Bias of Disease Prediction~\cite{zhao-etal-2024-llms}{,}\\ Racial and LGBTQ+ Biases~\cite{xie-etal-2024-addressing}{,} Out-of-Context Prompting~\cite{cotta2024out}{,} Attribute Neutral Modeling~\cite{hu2024enhancing}{,}\\ Personalized Alignment Techniques~\cite{kirk2024benefits}{,} Evaluating Biases in Context-Dependent~\cite{patel2024evaluating}}, leaf, text width=66em
                                 ]      
                        ]
                        [
                        \textbf{Evaluation and Detection Methods}, fill=blue!10
                                 [
                                {Evaluation Study~\cite{zack2024assessing}{,} BiasMedQA~\cite{schmidgall2024evaluation}{,} Hurtful Words~\cite{zhang2020hurtful}{,} Race-based Medicine~\cite{omiye2023large}{,}\\ Detect Debunked Stereotypes~\cite{swaminathan2024feasibility}{,} EquityMedQA~\cite{pfohl2024toolbox}{,} Superficial Fairness Alignment~\cite{wei2024actions}{,}\\ Examines Biased AI~\cite{adam2022just}{,} Identify Biases~\cite{Yang_2024}{,} Quantifying Cognitive Biases~\cite{lin-ng-2023-mind}{,}\\ Biases in Biomedical MLM~\cite{kim-etal-2023-race}{,} Bias of Disease Prediction~\cite{zhao-etal-2024-llms}{,} Racial and LGBTQ+ Biases~\cite{xie-etal-2024-addressing}{,}\\ FairMedFM~\cite{jin2024fairmedfm}}, leaf, text width=66em
                                 ]       
                        ]       
                        ]
                         [
                        \textbf{Explanability}, fill=blue!10
                                 [
                        \textbf{Benchmarks}, fill=blue!10
                                 [
                                    {FaReBio~\cite{fang-etal-2024-understanding}{,} Pathway2Text~\cite{yang-etal-2022-pathway2text}},  leaf, text width=66em  
                                 ]      
                        ]
                        [
                        \textbf{Enhance Methods}, fill=blue!10
                                 [
                                {Knowledge Graphs~\cite{shariatmadari2024harnessing}{,}  Medical Imaging Explainability~\cite{ghosh2023bridging}{,}  MedExQA~\cite{kim-etal-2024-medexqa}{,}\\ Retrieval and Reasoning on KGs~\cite{ji-etal-2024-retrieval}{,} DDCoT~\cite{zheng2023ddcot}{,} A ChatGPT Aided Explainable Framework~\cite{liu2023a}{,}\\ Medical Concept-Driven Attention~\cite{wang-etal-2022-novel}{,} FaReBio~\cite{fang-etal-2024-understanding}{,}\\ LLM-GCE~\cite{he-etal-2024-explaining}{,} kNN-Graph2Text~\cite{yang-etal-2022-pathway2text}{,} RAG-IM~\cite{mahbub2024from}{,} MedThink~\cite{gai2025medthink}},  leaf, text width=66em
                                 ]       
                        ]       
                        ]
                        ]
        \end{forest}
        }
    \caption{Summary of the recent research across various dimensions of trustworthiness of LLMs in healthcare. }
    \label{fig:Dimensions}
\end{figure*}

\subsection{Datasets}
The datasets used in studies of trust in LLMs for healthcare are categorized by the dimensions of trustworthiness they address in Appendix~\ref{app:datasets}, where we highlight key details such as data type, content, task, and dimensions of trustworthiness. The content of each dataset specifies its composition, while the task refers to the main problem to be solved for which the dataset is utilized. The data type varies across studies and includes web-scraped data, curated domain-specific datasets, public text corpora, synthetic data, real-world data, and private datasets, providing a comprehensive overview of their relevance to healthcare applications. 

Figure \ref{fig:datasetsTask} shows the number of studies using three major dataset types—Med-QA (blue), Med-Gen (orange), and Med-IE (green)—in relation to six trustworthiness dimensions: Truthfulness, Privacy, Safety, Robustness, Fairness and Bias, and Explainability. Figure \ref{fig:datasetsTask} shows how three major dataset types—Med-QA, Med-Gen, and Med-IE—are used across six trust dimensions. Truthfulness is most studied with both Med-QA and Med-Gen. Med-QA is also common in fairness and explainability, while Med-Gen contributes to safety and privacy. Med-IE, though less used overall, is more prominent in robustness and explainability. This highlights the dominance of Med-QA and Med-Gen, with Med-IE offering value in specific areas of trustworthiness.

\subsection{Models}
The models assessed in studies on trust in LLMs for the healthcare domain are outlined, along with their trustworthiness dimensions, in Appendix~\ref{app:gpts}, where we summarized key details such as the model name, release year, openness, architecture, task, and the institution responsible for its development. Figure \ref{fig:modelsDimensions} illustrates the proportions of different model types—open-source, closed-source, and architectures including encoder-only, decoder-only, and encoder-decoder—used in research addressing various trustworthiness aspects of LLMs in healthcare: Explainability, Fairness and Bias, Robustness, Safety, Privacy, and Truthfulness. From Figure \ref{fig:modelsDimensions}, it is clear that Decoder-only and Open-source models are the most commonly used across all trustworthiness dimensions—especially in robustness, explainability, and truthfulness—highlighting their accessibility and alignment with generative tasks. Closed-source models appear more in fairness and privacy studies, while Encoder-only and Encoder-decoder models are used less frequently, mostly in fairness and truthfulness evaluations. 

\subsection{Tasks}
The tasks covered various primary focuses of LLMs in healthcare. Inspired from the survey by \citet{liu2024survey}, these tasks include:

\paragraph{Medical Information Extraction (Med-IE)} Med-IE extracts structured medical data from unstructured sources such as EHRs, clinical notes, and research articles. Key tasks include entity recognition (identifying diseases, symptoms, and treatments), relationship extraction (understanding entity connections), event extraction (detecting clinical events and attributes), information summarization (condensing medical records), and adverse drug event detection (identifying medication-related risks).

\paragraph{Medical Question Answering (Med-QA)} Med-QA systems interpret and respond to complex medical queries from patients, clinicians, and researchers. Their core functions include query understanding (interpreting user questions), information retrieval (finding relevant data in medical databases), and inference and reasoning (drawing conclusions, inferring relationships, and predicting outcomes based on retrieved data).

\paragraph{Medical Natural Language Inference (Med-NLI)} Med-NLI analyzes the logical relationships between medical texts. Key tasks include textual entailment (determining if one statement logically follows another), contradiction detection (identifying conflicting statements), neutral relationship identification (recognizing unrelated statements), and causality recognition (inferring cause-and-effect relationships).

\paragraph{Medical Text Generation (Med-Gen)} Med-Gen focuses on generating and summarizing medical content. Its key applications include text summarization (condensing lengthy documents into concise summaries) and content generation (producing new medical descriptions or knowledge based on input data).

\section{Trustworthiness of LLMs in Healthcare}
We examine the challenges related to the trustworthiness of LLMs in healthcare, outlining key strategies for identifying and mitigating these concerns. From our literature review screening, we identified truthfulness, privacy, safety, robustness, fairness and bias, and explainability as key trustworthiness dimensions of LLMs as highlighted in TrustLLM \cite{huang2024position}, particularly in healthcare. Figure \ref{fig:Dimensions} provides a summary of the recent research on trust in LLMs for healthcare across key dimensions of trustworthiness.

\subsection{Truthfulness}
\begin{tcolorbox}[colback=cyan!10, colframe=cyan!50!black,
title=Findings in Truthfulness,
fonttitle=\bfseries,
coltitle=black,
boxrule=0.6mm,
arc=2mm,
left=2mm, right=2mm, top=1mm, bottom=1mm]
Self-reflection and fact-checking reduce hallucinations but do not scale or generalize well, especially for long-form clinical contexts. Improving truthfulness will require hybrid pipelines that combine retrieval, reasoning, multi-agent self-correction, and uncertainty estimation.
\end{tcolorbox}
Ensuring the \emph{truthfulness} of LLMs in healthcare is critical, as inaccurate or fabricated information can directly harm clinical decisions. Hallucinations arise from biased data, weak contextual reasoning, and unverifiable sources \cite{ahmad2023creating}. Current work targets factual reliability via benchmarking, post-hoc correction, uncertainty quantification, and improved evidence synthesis.

Several benchmarks have emerged to quantify and categorize hallucinations. The Med-HALT benchmark \cite{pal-etal-2023-med} evaluates hallucination types using reasoning-based tests (e.g., “False Confidence”) and memory checks. In multimodal settings, Med-HVL \cite{yan2024med} distinguishes between Object Hallucination and Domain Knowledge Hallucination.

To mitigate hallucinations, post-hoc correction techniques are gaining traction. MEDAL \cite{li-etal-2024-better} presents a model-agnostic self-correction module that improves summarization outputs without retraining. Similarly, interactive feedback strategies like self-reflection loops \cite{ji-etal-2023-towards} allow LLMs to iteratively refine their responses.

Uncertainty quantification approaches provide complementary detection tools. \citet{farquhar2024detecting} apply semantic entropy to flag low-confidence responses, while SEPs \cite{han2024semantic} offer a lightweight, hidden-state-based approximation suited for clinical use.

Recent efforts also examine the trustworthiness of evidence synthesis pipelines. \citet{zhang2024leveraging} highlight risks when LLMs generate clinical summaries without grounding, emphasizing the need for transparency in literature retrieval and evidence aggregation. Debate-based evaluation, as explored in MAD \cite{smit2023we}, introduces multi-agent deliberation to vet factual consistency in medical QA. Finally, SEND \cite{mohammadzadeh2024hallucination} introduces a neuron dropout technique to detoxify hallucination-prone neurons during training, aiming to improve inherent model truthfulness.

Factual accuracy is critical for trust in healthcare LLMs, where clinical safety relies on reliable, verifiable outputs. Yet, current models often produce ungrounded content and lack source traceability. Recent work addresses this through medical claim benchmarks, self-correction, automated fact-checking, multi-turn verification, and multi-perspective reasoning—advancing transparency, factuality, and clinical relevance.

To support systematic validation, \citet{akhtar-etal-2022-pubhealthtab} introduce PubHealthTab, a table-based dataset for checking public health claims against noisy evidence, while \citet{sarrouti-etal-2021-evidence-based} propose HEALTHVER, a benchmark for evidence-based fact-checking tailored to medical claims. These resources enable structured evaluation of LLM outputs and form the foundation for improving medical claim verification.

Beyond static benchmarks, dynamic self-correction methods have shown promise. \citet{gou2024critic} propose CRITIC, a framework inspired by human fact-checking, in which LLMs iteratively assess and revise their own responses. This process mimics expert reasoning and introduces a layer of critical reflection into model outputs. Complementing this, \citet{cohen2023lm} present a cross-examination approach, where a second "examiner" model engages in multi-turn dialogue to probe for factual inconsistencies in the original response. While CRITIC emphasizes human-like evaluation, cross-examination leverages interaction between models to simulate external verification.

To further reduce hallucinations and improve factual consistency, \citet{tan2024faithful} introduce a method that incorporates multiple scientific perspectives when resolving conflicting arguments, strengthening LLMs’ reasoning capabilities through broader contextual understanding.

\paragraph{Evaluations} Truthfulness is assessed with hallucination/factuality benchmarks (e.g., Med-HALT) and feedback-loop strategies; expert annotations on HealthSearchQA, MedQA, and MultiMedQA are common but costly and subjective.

\paragraph{Limitations} Evaluations remain fragmented: narrow task coverage, varying definitions, closed-source dependencies, and limited generalizability across modalities/clinical domains. Many self-correction methods are task-specific and lack robustness.

\subsection{Privacy}
\begin{tcolorbox}[colback=cyan!10, colframe=cyan!50!black,
title=Findings in Privacy,
fonttitle=\bfseries,
coltitle=black,
boxrule=0.6mm,
arc=2mm,
left=2mm, right=2mm, top=1mm, bottom=1mm]
LLMs pose serious privacy risks from memorizing and regenerating PHI. Differential privacy and federated learning help but often hurt utility; future work needs fine-grained, instance-level risk estimation across training and inference.
\end{tcolorbox} 
LLMs in healthcare face end-to-end \emph{privacy} risks due to their tendency to memorize and potentially regenerate sensitive data such as protected health information (PHI) \cite{das2024security, pan2020privacy}. Key threats include data memorization, insufficient de-identification, and the privacy-utility trade-offs of fine-tuning methods. This section examines current vulnerabilities, mitigation strategies, and emerging approaches for achieving privacy-preserving healthcare LLMs.

Data memorization is a core concern, especially in domain-specific models like Medalpaca \cite{han2025medalpacaopensourcecollection}, which are more likely to retain PHI and pose heightened re-identification risks \cite{yang2024memorization}. Structured attacks like those demonstrated in SecureSQL \cite{song-etal-2024-securesql} reveal that even chain-of-thought (CoT) prompting provides only marginal defense against leakage.

Pre-training privacy measures include de-identification techniques like GPT-4 masking \cite{liu2023deid} and synthetic note generation \cite{altalla2025evaluating}, though these offer limited protection. \citet{xin2024a} caution that such methods may create a false sense of security, as subtle semantic cues can still lead to PHI leakage.

Fine-tuning methods such as federated learning \cite{zhao2024llm} and differential privacy (DP) \cite{singh2024whispered} provide stronger safeguards by decentralizing data or adding noise to protect individual records. However, these methods often compromise model performance or scalability \cite{liu2024survey}.

Emerging techniques seek to reduce this trade-off. APNEAP \cite{wu-etal-2024-mitigating-privacy} introduces activation patching for privacy neuron editing, reducing leakage without harming utility. Complementarily, \citet{chen2024generative} offer a broader survey of privacy risks and solutions across generative AI use cases in healthcare.

Ethical and personalization challenges further complicate privacy design. \citet{zhui2024ethical} emphasize building privacy-conscious frameworks in medical education, while \citet{kirk2024benefits} caution that overly personalized alignment strategies may inadvertently violate user privacy, advocating instead for bounded personalization.

\paragraph{Evaluations} Typical assessments use membership/attribute inference and reconstruction attacks, plus privacy–utility analyses (e.g., redaction or DP) under controlled settings. Real-world audits are scarce, and proposed risk-prediction or memorization-severity frameworks lack medical-specific benchmarks.

\paragraph{Limitations} Current defenses (e.g., DP, redaction) still trade performance for privacy. Many evaluations assume idealized adversaries, and systematic ways to balance memorization risk and utility—especially for multimodal, long-context models—remain limited.

\subsection{Safety}
\begin{tcolorbox}[colback=cyan!10, colframe=cyan!50!black,
title=Findings in Safety,
fonttitle=\bfseries,
coltitle=black,
boxrule=0.6mm,
arc=2mm,
left=2mm, right=2mm, top=1mm, bottom=1mm]
Medical LLMs can still produce harmful or misleading content after safety tuning. Benchmarks expose vulnerabilities to adversarial prompts and embedded misinformation. Robust safety demands proactive alignment and multi-stage, clinically grounded evaluations that simulate realistic misuse.
\end{tcolorbox}
Ensuring \emph{safety} is critical: small weight edits ($\sim1.1\%$) can implant lasting biomedical falsehoods without hurting average performance \cite{han2024medical}, and poisoning as little as 0.001\% of training data can embed persistent misinformation \cite{han2024dataPoisoning}. Key concerns include the ease of injecting persistent falsehoods into model weights, inadequate performance on harmful prompts, trade-offs between safety alignment and hallucination, and privacy-related vulnerabilities that can escalate safety risks. This section explores current benchmarks, safety alignment strategies, and the overlap between safety and privacy threats.

To systematically evaluate harmful outputs, benchmarks like MedSafetyBench \cite{han2024medsafetybench} and Med-Harm \cite{han2024towards} use adversarial and real-world queries to test model responses. Results show that even medically fine-tuned LLMs often fail safety criteria unless specifically optimized. MEDIC \cite{kanithi2024mediccomprehensiveframeworkevaluating} broadens this evaluation across dimensions such as reasoning and reliability, offering a holistic safety diagnostic tool.

Safety alignment remains challenging due to its tension with other objectives. UNIWIZ couples safety-driven training with retrieval to reduce unsafe outputs while preserving accuracy \cite{das2024uniwiz}. However, over-alignment increases hallucination, whereas under-alignment permits unsafe behavior, demonstrating the delicate balance required for clinical reliability.

Finally, privacy threats intersect with safety risks. \citet{leemann2024is} show that membership inference attacks, like Gradient Likelihood Ratio (GLiR), can detect whether individual patient data was used in training. This not only violates privacy but also raises safety concerns, as misuse of sensitive information can misguide clinical outcomes.

\paragraph{Evaluations} Methods include manual red teaming, automated stress tests, and healthcare-specific tasks (e.g., drug interactions, diagnostic advice) with expert review, though many prompts still derive from general domains.

\paragraph{Limitations} Mitigations often target generic harms rather than medical-specific risks (e.g., dangerous dosing). Red teaming rarely involves medical experts, and clinically grounded benchmarks with expert-in-the-loop validation remain limited.

\subsection{Robustness}
\begin{tcolorbox}[colback=cyan!10, colframe=cyan!50!black,
title=Findings in Robustness,
fonttitle=\bfseries,
coltitle=black,
boxrule=0.6mm,
arc=2mm,
left=2mm, right=2mm, top=1mm, bottom=1mm]
LLMs are fragile under distribution shifts, adversarial prompts, and instruction changes. Existing defenses (adversarial testing, test-time adaptation) are often task-specific. Robustness demands context-aware evaluation, multi-agent training, and resilience to real-world perturbations.
\end{tcolorbox}
Ensuring \emph{robustness} is vital for clinical deployment. Challenges include adversarial vulnerability, sensitivity to domain shifts and instruction variations, and prompt-based attacks. To address these issues, recent work explores adversarial testing, test-time adaptation, prompt security, data augmentation, and instruction robustness strategies.

Adversarial robustness is addressed through synthetic data generation. \citet{yuan2023revisiting} and \citet{wang2024stumbling} introduce adversarial test samples tailored to the medical domain, such as synthetic anomaly cases and boundary stress testing, to assess model resilience. \citet{alberts2023large} emphasize the importance of aligning adversarial testing methods with real-world medical complexities. In parallel, \citet{gallifant-etal-2024-language} reveal that simply substituting generic and brand drug names within biomedical benchmarks leads to performance drops of up to 10\%, highlighting the fragility of LLMs to clinically trivial lexical shifts.

Uncertainty quantification offers another avenue for robustness. LLM-TTA \cite{o2024improving} explores test-time adaptation techniques to enhance model performance on rare or unfamiliar cases, common in medical diagnostics. This approach complements adversarial robustness by identifying instances where models are likely to err.

Instruction robustness is examined by \citet{ceballos-arroyo-etal-2024-open}, who find that specialized medical models may be more fragile than general-purpose models when instructions are reworded, suggesting that excessive domain adaptation may reduce flexibility.

Prompt security is enhanced by \citet{tang-etal-2024-secure}, who introduce a framework that strengthens LLM robustness with cryptographic prompt authentication, mitigating vulnerabilities associated with prompt injections and adversarial attacks.

Data augmentation techniques are employed in MEDSAGE \cite{binici2025medsage}, which uses LLM-generated synthetic dialogues to simulate ASR errors, improving the robustness of medical dialogue summarization systems. Similarly, RobustQA \cite{han-etal-2023-robustqa} benchmarks the robustness of domain adaptation for open-domain question answering across diverse domains, facilitating the evaluation of ODQA’s domain robustness.

Lastly, prompt engineering strategies, such as out-of-context prompting, are explored by \citet{cotta2024out}, who demonstrate that applying random counterfactual transformations can improve the fairness and robustness of LLM predictions without additional data or fine-tuning.

\paragraph{Evaluations} Robustness is tested via distribution shifts, adversarial or out-of-domain inputs, synthetic perturbations, and black-box stress tests. Newer setups (e.g., MedQA-Adversarial, RAG robustness tests) probe noisy retrieval and unfamiliar conditions, but clinical realism and standardization remain limited.

\paragraph{Limitations} Lack of standardized, clinically grounded stress tests; overreliance on synthetic or narrow tasks; and brittle/costly mitigations (e.g., adversarial training, RAG) impede deployment. Multimodal and long-context robustness remain underexplored.

\subsection{Fairness and Bias}
\begin{tcolorbox}[colback=cyan!10, colframe=cyan!50!black,
title=Findings in Fairness,
fonttitle=\bfseries,
coltitle=black,
boxrule=0.6mm,
arc=2mm,
left=2mm, right=2mm, top=1mm, bottom=1mm]
Bias across race, gender, and identity persists in medical LLMs. New benchmarks and mitigations help but are often narrow or misaligned with clinical reality. Progress requires intersectional audits, inclusive datasets, and collaboration with affected communities.
\end{tcolorbox}
Ensuring \emph{fairness} is essential: biased outputs can exacerbate disparities in access, diagnosis, and treatment. Key areas of concern include demographic bias (e.g., race, gender, identity), automated detection of these biases, mitigation strategies based on model accessibility, and the need for ethical clarity and conceptual frameworks. Recent work spans benchmark creation, debiasing techniques, prompt interventions, and calls for more transparent fairness evaluations.

Bias identification remains a foundational step. Studies show that LLMs can replicate and even amplify racial, gender, and identity-based biases. For example, \citet{omiye2023large}, \citet{zack2024assessing}, and \citet{kim-etal-2023-race} highlight persistent demographic biases in medical responses. \citet{zhao-etal-2024-llms} find that diagnostic recommendations vary unfairly by demographic group, while \citet{xie-etal-2024-addressing} reveal systematic inequities in outputs concerning race and LGBTQ+ identities. \citet{patel2024evaluating} further demonstrate that LLMs can reinforce social and gender-based stereotypes in sensitive areas such as sexual and reproductive health, underscoring the risks in context-dependent medical interactions.

Detection and benchmarking tools help quantify and monitor these disparities. \citet{swaminathan2024feasibility} propose tools for identifying race-based stereotypes in medical Q\&A. Benchmarks such as BiasMedQA \cite{schmidgall2024evaluation}, EquityMedQA \cite{pfohl2024toolbox}, and FairMedFM \cite{jin2024fairmedfm} offer frameworks for testing model behavior across diverse patient profiles and clinical contexts.

Mitigation strategies differ by model accessibility. For open-source models, techniques like adversarial debiasing \cite{zhang2020hurtful}, causal intervention (CI4MRC) \cite{zhu2023causal}, multi-agent collaboration \cite{ke2024mitigating}, and attribute-neutral modeling \cite{hu2024enhancing} are applied to reduce bias. Data augmentation \cite{parray2023chatgpt} and bias-aware embedding assessments \cite{lin-ng-2023-mind} provide further tools to enhance fairness in pretraining and inference.

Closed-source models present unique challenges due to limited transparency. In these cases, fairness is addressed via instruction fine-tuning \cite{singhal2023large}, external prompt engineering \cite{schmidgall2024evaluation}, or bounded personalization strategies \cite{kirk2024benefits}, though these are less interpretable and harder to audit.

Ethical and conceptual considerations also play a role. \citet{wei2024actions} call for distinguishing between intrinsic and behavioral fairness, while \citet{zhui2024ethical} and \citet{cotta2024out} promote fairness through education and prompt design. Finally, \citet{adam2022just} and \citet{Yang_2024} warn that unchecked bias can distort care decisions and patient trust, emphasizing the stakes of fairness in real-world applications.

\paragraph{Evaluations} Assess disparities across subgroups using quantitative metrics (performance gaps, stereotyping scores) and qualitative audits; new benchmarks (FairMedFM, EquityMedQA) target equity in clinical recommendations, though intersectional analyses remain limited.

\paragraph{Limitations} Many evaluations overlook intersectional/institutional biases, adapt generic NLP methods without clinical causal context, and lack longitudinal assessment. Integration with other trust dimensions (robustness, privacy) is limited.

\subsection{Explanability}
\begin{tcolorbox}[colback=cyan!10, colframe=cyan!50!black,
title= Findings in Explainability,
fonttitle=\bfseries,
coltitle=black,
boxrule=0.6mm,
arc=2mm,
left=2mm, right=2mm, top=1mm, bottom=1mm]
Despite advances in rationales and attention maps, most tools lack clinical relevance and faithfulness. Methods often misalign with clinician reasoning; progress needs domain-specific frameworks plus causal/counterfactual explanations.
\end{tcolorbox}
Lack of \emph{explainability} limits clinical trust. Recent research explores both intrinsic (model-integrated) and post-hoc (output-interpretation) techniques to make LLM reasoning more interpretable. These methods span a wide range of modalities, including text, graphs, tables, and images, and often incorporate domain-specific knowledge or human-centered reasoning to bridge model outputs and clinical expectations.

Intrinsic explainability methods enhance transparency by aligning model attention with medical knowledge. For example, \citet{shariatmadari2024harnessing} integrate knowledge graphs with attention visualization, while \citet{wang-etal-2022-novel} use Wikipedia-derived medical concepts to guide attention for code prediction, resulting in more concept-consistent outputs. Similarly, structure-to-text models like Pathway2Text \cite{yang-etal-2022-pathway2text} convert biomedical graphs into interpretable narratives, supporting a more intuitive understanding of complex structured inputs.

Post-hoc strategies focus on generating faithful rationales and justifications. FaReBio \cite{fang-etal-2024-understanding} highlights how summarization faithfulness suffers with increased abstractiveness and introduces a benchmark to evaluate reasoning fidelity. In the molecular domain, LLM-GCE \cite{he-etal-2024-explaining} generates counterfactuals for Graph Neural Networks (GNNs) using dynamic feedback to ensure chemically valid, interpretable explanations.

Several methods target zero-shot interpretability without task-specific fine-tuning. RAG-IM \cite{mahbub2024from} enables table-based clinical predictions with natural language justifications, while \citet{liu2023a} embed ChatGPT into a diagnostic workflow with integrated interpretability components. Retrieval-based systems such as Retrieval + KG \cite{ji-etal-2024-retrieval} and DDCoT \cite{zheng2023ddcot} further enhance reasoning by chaining knowledge-grounded prompts across modalities.

Explainability in imaging and multimodal contexts is also gaining traction. MedThink \cite{gai2025medthink} fuses visual and textual inputs to improve multimodal reasoning, and MedExQA \cite{kim-etal-2024-medexqa} supplies detailed rationales for visual question answering. \citet{ghosh2023bridging} decomposes black-box decisions into expert modules with first-order logic (FOL) reasoning.

\paragraph{Evaluations} Assessments use attribution heatmaps (e.g., LIME/SHAP), human-in-the-loop ratings, and contrastive/instruction-following tests focused on clarity, factual alignment, and clinical usefulness; standardized healthcare benchmarks remain scarce.

\paragraph{Limitations} Attribution tools have uncertain clinical validity, few studies show improved clinician decisions, and claims are rarely compared across models. Many methods don’t scale to large, multimodal, instruction-following LLMs, leaving faithfulness and practicality unresolved.

\subsection{Cost and Efficiency Considerations}
While large language models offer transformative potential for healthcare, their real-world deployment faces substantial cost and efficiency constraints. Large medical LLMs (e.g., GPT-4, Med-PaLM 2) are expensive to train, fine-tune, and operate, requiring significant compute, memory, and HIPAA-compliant infrastructure; latency and resource demands further hinder use in low-resource settings. These constraints restrict access to well-funded institutions and slow real-world adoption. Smaller open-source models lower inference costs and enable local/edge deployment but often reduce performance—especially in truthfulness, safety, and robustness. Scaling AI in healthcare therefore requires balancing trustworthiness with computational efficiency.

\section{Future Directions}
While core trust dimensions, truthfulness, privacy, robustness, fairness, explainability, and safety, have been the focus of recent work, emerging model paradigms such as multi-agent systems, multi-modal models, and small open-source LLMs introduce new trust challenges underexplored.

\paragraph{Multi-Agent LLMs} Multi-agent LLMs enable distributed reasoning through collaboration between specialized agents, offering improved robustness and self-correction. However, they also raise concerns around coordination, error propagation, and the interpretability of inter-agent communication. Trustworthy multi-agent systems will require protocols for communication, verification, and evaluation that ensure factual alignment and fairness. For example, \citet{lu-etal-2024-triageagent} introduce TriageAgent, a clinical multi-agent framework with role-specific LLMs for diagnosis and decision-making. While it shows benefits like structured collaboration and early stopping, it also reveals trust challenges, including inconsistent agent confidence, limited transparency, and error propagation—highlighting the need for stronger verification and alignment in high-stakes settings.

\paragraph{Multimodal Foundation Models} Multi-modal LLMs combine text, images, and structured data, better reflecting real-world clinical inputs but complicating trust evaluation. Challenges include cross-modal hallucination, misalignment, and reduced explainability. Addressing these issues will require modality-specific assessments, interpretable fusion strategies, and fairness testing across both textual and visual modalities. For example, \citet{liu-etal-2024-geneverse} evaluate open-source multimodal LLMs for genomics and proteomics, highlighting issues with factual consistency and alignment across modalities—underscoring the importance of structured evaluation and interpretable model design in biomedical contexts.

\paragraph{Small Open-Source LLMs} Small open-source medical LLMs are gaining traction for their transparency, adaptability, and lower computational demands, making them attractive for deployment in resource-constrained or privacy-sensitive settings. However, their reduced capacity often leads to increased hallucinations, weaker safety alignment, and heightened privacy risks during fine-tuning on limited clinical data. Ensuring their trustworthiness requires lightweight hallucination mitigation, privacy-preserving training, and scalable evaluation pipelines. Despite their growing use, few studies directly examine these trust issues in small medical LLMs, as most existing research focuses on larger or general-purpose models, leaving a critical gap in the literature.

\section{Conclusion}
As large language models continue to expand their role in healthcare, ensuring their trustworthiness remains a critical challenge. This survey reviewed six core dimensions—truthfulness, privacy, safety, robustness, fairness, and explainability—highlighting key methods, benchmarks, and limitations in current research. While recent advances have laid important groundwork, most existing solutions remain narrowly scoped and lack integration across dimensions, limiting their effectiveness in real-world clinical settings.

\section*{Acknowledgment}
Our work is sponsored by NSF \#2442253, NAIRR Pilot with PSC Neocortex and NCSA Delta, Commonwealth Cyber Initiative, Children’s National Hospital, Fralin Biomedical Research Institute (Virginia Tech), Sanghani Center for AI and Data Analytics (Virginia Tech), Virginia Tech Innovation Campus, and generous gifts from Nivida, Cisco, and the Amazon + Virginia Tech Center for Efficient and Robust Machine Learning.

\clearpage
\section*{Limitations}
This survey provides a comprehensive overview of the challenges associated with LLMs in healthcare, but it primarily focuses on existing methodologies, leaving out emerging technologies that could address these issues in new ways. It also lacks practical insights into the real-world implementation of these solutions, such as deployment challenges, cost considerations, and system integration, which would make the findings more applicable to healthcare settings.

While the paper addresses privacy and safety, it does not fully explore broader ethical issues like informed consent, patient autonomy, and human oversight. Additionally, the survey focuses on current research without delving into the long-term societal and health impacts of LLM deployment, such as changes in doctor-patient relationships, patient trust, and healthcare workflows.
% Bibliography entries for the entire Anthology, followed by custom entries
%\bibliography{anthology,custom}
% Custom bibliography entries only

\bibliography{acl_latex}

\clearpage
\appendix

\section{Inclusion \& Exclusion Criteria Details}
\label{app:InclusionExclusionCriteriaDetails}
We conducted an extensive search to identify peer-reviewed papers that address the trustworthiness of LLMs in the healthcare domain. Our search strategy involved a wide range of keyword combinations related to LLMs and core trust dimensions, including: trustworthiness, truthfulness, privacy, safety, robustness, fairness, bias, and explainability. We targeted both domain-specific and general AI venues, focusing on recent publications from 2022 onward.

Specifically, we searched across top-tier conferences and journals, including ACL, EMNLP, NAACL, ICML, NeurIPS, ICLR, KDD, AAAI, IJCAI, Nature, and Science, using platforms such as Google Scholar, Nature, and Science. A full list of keyword queries used in our search is provided below. These queries combined domain terms (medical, clinical) with trust-related dimensions, applied across both “large language models” and “foundation models.” Examples include:
\begin{itemize}[leftmargin=*]
\item large language models, medical, explainability  
\item large language models, medical, explainable 
\item foundation model, medical, explainability 
\item large language models, clinical, explainability 
\item large language models, medical, truthfulness  
\item large language models, medical, trustworthiness 
\item foundation model, medical, trustworthiness 
\item large language models, clinical, truthfulness 
\item large language models, clinical, safety 
\item large language models, medical, safety  
\item foundation model, medical, safety 
\item large language models, clinical, fairness 
\item large language models, medical, fairness
\item foundation model, medical, fairness 
\item large language models, clinical, robustness 
\item foundation model, medical, robustness 
\item large language models, medical, robustness 
\item large language models, clinical, privacy 
\item large language models, medical, privacy  
\item foundation model, medical, privacy 
\item large language models, clinical, ethics 
\item large language models, medical, ethics 
\item foundation model, medical, ethics 
\end{itemize}

In total, our initial search returned approximately 15,322 results, including duplicates and non-relevant papers. Our filtering process proceeded in three stages:
\begin{itemize}
\item Duplicate removal – approximately 11,172 papers eliminated.
\item Relevance screening – we excluded papers that:
(a) did not focus on trustworthiness aspects (e.g., architecture design or multi-modal fusion techniques),
(b) were not specific to the healthcare domain, or
(c) were unpublished preprints (e.g., arXiv manuscripts).
\item Final selection – we curated a final set of 62 papers that directly addressed trust-related challenges in healthcare LLMs, focusing on one or more of the following dimensions: truthfulness, privacy, safety, robustness, fairness, bias, and explainability.
\end{itemize}
\clearpage
\section{Comparison of Datasets}
\label{app:datasets}
We systematically collected and analyzed 38 datasets relevant to the study of trust in LLMs for healthcare. Table 1 provides a comprehensive summary, highlighting key attributes such as data type, content, associated tasks, and the specific trustworthiness dimensions they address. These datasets vary widely, including web-scraped data, curated domain-specific datasets, public text corpora, synthetic data, real-world data, and private datasets. Each dataset's content specifies its composition, while its associated task defines its primary research application. Additionally, we categorize the datasets based on critical trustworthiness dimensions—truthfulness, privacy and safety, robustness, fairness and bias, and explainability—offering a structured evaluation of their contributions to building reliable and trustworthy healthcare AI.

\begin{table*}[htbp]
        \centering
        \small
        \renewcommand{\arraystretch}{1.2}
        \setlength{\tabcolsep}{6pt}
        \begin{tabular}{p{2cm} p{2.5cm} p{4cm} p{3.5cm} p{1.5cm}}
            \toprule
            \textbf{Datasets} & \textbf{Data Type} & \textbf{Content} & \textbf{Task} & \textbf{Dimensions} \\
            \midrule
        \href{https://huggingface.co/datasets/openlifescienceai/multimedqa}{MultiMedQA} &
          Combination of Public and Synthetic Data, Curated Domain-Specific Dataset &
          208,000 entries. A benchmark combining six existing medical questions answering datasets spanning professional medicine, research and consumer queries and a new dataset of medical questions searched online, HealthSearchQA. &
          (Med-QA) Tasks including Medical Question Answering, Clinical Reasoning, Evidence-Based Medicine, Multilingual and Multimodal Support, Bias and Safety Analysis &
          Fairness and Bias \\
          \hline
            \href{https://pure.johnshopkins.edu/en/publications/evaluation-and-mitigation-of-cognitive-biases-in-medical-language}{BiasMedQA} &
          Curated Domain-Specific Datasets &
          1273 USMLE questions &
          (Med-QA) Replicate common clinically relevant cognitive biases &
          Fairness and Bias \\
          \hline
            \href{https://arxiv.org/abs/2403.12025?utm_source=chatgpt.com}{EquityMedQA} &
          Curated domain-specific datasets and synthetic data &
          4,619 examples. Cover a wide range of medical topics to surface biases that could harm health equity, including implicit and explicit adversarial questions addressing biases like stereotypes, lack of structural explanations, and withholding information. &
          (Med-QA) Evaluate the performance of LLMs in generating unbiased, equitable medical responses. &
          Fairness and Bias \\
          \hline
            \href{https://rajpurkar.github.io/SQuAD-explorer/}{SQuAD} &
          Curated Domain-Specific Dataset &
          Consists of over 100,000 question-answer pairs derived from more than 500 articles from Wikipedia. Each question is paired with a segment of text from the corresponding article, serving as the answer. &
          (Med-QA)To develop models that can read a passage and answer questions about it, assessing the model's ability to understand and extract information from the text. &
          Fairness and Bias \\
          \hline
            \href{https://www.nature.com/articles/sdata201635}{MIMIC- III} &
          Public text corpora, real-world data &
          De-identified health-related data from over 40,000 critical care patients, including demographics, vital signs, laboratory tests, medications, and caregiver notes. &
          (Med-IE) Epidemiological studies, clinical decision-rule improvement, machine learning in healthcare. &
          Fairness and Bias, Explainability, Robustness \\
          \hline
            \href{https://github.com/jind11/MedQA}{MedQA} &
          Curated Domain-Specific Datasets &
          194,000 multiple-choice medical exam questions. A benchmark that includes questions drawn from the United States Medical License Exam (USMLE). &
          (Med-QA) Exam the physicians to test their ability to make clinical decisions &
          Fairness and Bias, Robustness, Explainability, Truthfulness, Privacy \\
          \hline
            \href{https://www.nature.com/articles/s41597-023-02814-8}{PMC-Patients} &
          Curated dataset derived from public text corpora. &
          Contains 167,000 patient summaries extracted from 141,000 PMC articles &
          (Med-IE) Designed to benchmark ReCDS systems through two primary tasks: Patient-to-Article Retrieval (PAR), Patient-to-Patient Retrieval (PPR) &
          Robustness \\
          \hline
            \href{(https://arxiv.org/html/2403.03744v4)}{MedSafetyBench} &
          Curated domain-specific dataset and synthetic (generated using GPT-4, Llama-2-7b-chat, and adversarial techniques). &
          1,800 harmful medical requests violating medical ethics, along with 900 corresponding safe responses. The dataset is structured based on the Principles of Medical Ethics from the American Medical Association (AMA). &
          (Med-Gen) Assess the medical safety of LLMs by testing whether they refuse to comply with harmful medical requests. Fine-tune LLMs using medical safety demonstrations to enhance their alignment with ethical medical guidelines. &
          Safety \\
          \hline
            \bottomrule
        \end{tabular}
        \label{tab:Datasets}\end{table*}

\begin{table*}[htbp]
        \centering
        \small
        \renewcommand{\arraystretch}{1.2}
        \setlength{\tabcolsep}{6pt}
        \begin{tabular}{p{2cm} p{2.5cm} p{4cm} p{3.5cm} p{1.5cm}}
            \toprule
            \textbf{Datasets} & \textbf{Data Type} & \textbf{Content} & \textbf{Task} & \textbf{Dimensions} \\
            \midrule
        \href{https://aclanthology.org/2024.findings-acl.102.pdf}{UNIWIZ} &
          Synthetic and curated data, including: 17,638 quality-controlled conversations, and 10,000 augmented preference data &
          17,638 conversations and 10,000 augmented preference data. Features conversations that integrate safety and knowledge alignment. A "safety-priming" method was employed to generate synthetic safety data, and factual information was injected into conversations by retrieving content from curated sources. &
          (Med-Gen) Fine-tune large language models to enhance their performance in generating safe and knowledge-grounded conversations. &
          Safety \\
          \hline
            \href{https://github.com/allenai/scifact?utm_source=chatgpt.com}{SciFact} &
          Curated Domain-Specific Dataset. &
          2,011 claims. Includes claims and corresponding evidence abstracts, each annotated with labels indicating whether the claim is supported or refuted, along with rationales justifying the decision. &
          (Med-Gen) To verify the veracity of scientific claims by identifying supporting or refuting evidence within abstracts and providing justifications for these decisions. &
          Truthfulness \\
          \hline
            \href{https://aclanthology.org/2022.findings-naacl.1/?utm_source=chatgpt.com}{PubHealthTab} &
          Curated Domain-Specific Dataset &
          Contains 1,942 real-world public health claims, each paired with evidence tables extracted from over 300 websites. &
          (Med-Gen) Facilitates evidence-based fact-checking by providing claims and corresponding evidence tables for verification. &
          Truthfulness \\
          \hline
            \href{https://github.com/facebookresearch/LAMA?utm_source=chatgpt.com}{LAMA} &
          Curated Domain-Specific Dataset. &
          24,223 entries of knowledge sources. Comprises a set of knowledge sources, each containing a collection of facts. &
          (Med-Gen) To probe pretrained language models to determine the extent of their factual and commonsense knowledge. &
          Truthfulness \\
          \hline
            \href{https://aclanthology.org/P17-1147/}{TriviaQA} &
          Curated Domain-Specific Dataset. &
          Consists of over 650,000 question-answer pairs, each linked to a set of supporting documents. The questions are sourced from trivia websites, and the answers are derived from the corresponding documents. &
          (Med-QA) Training and evaluating models on reading comprehension, specifically focusing on the ability to extract and reason over information from provided documents to answer questions. &
          Truthfulness \\
          \hline
            \href{https://aclanthology.org/Q19-1026/}{Natural Questions (NQ)} &
          Real data &
          ~99.80 GB, with downloaded files accounting for 45.07 GB and the generated dataset occupying 54.73 GB. consists of real anonymized queries from Google's search engine users, paired with answers derived from entire Wikipedia articles. &
          (Med-QA) To develop and evaluate question-answering systems that can read and comprehend entire Wikipedia articles to find answers to user queries. &
          Truthfulness \\
          \hline
            \href{https://arxiv.org/abs/2212.10511}{PopQA} &
          Curated Domain-Specific Dataset. &
          consists of 14,000 QA pairs, each associated with fine-grained Wikidata entity IDs, Wikipedia page views, and relationship type information. &
          (Med-QA) Designed for open-domain question answering tasks, focusing on evaluating the effectiveness of language models in retrieving and utilizing factual knowledge. &
          Truthfulness \\
          \hline
            \href{https://fever.ai/dataset/fever.html}{FEVER} &
          Curated Domain-Specific Dataset. &
          comprises 185,000 claims, each paired with evidence from Wikipedia articles. These claims are categorized as supported, refuted, or not verifiable. &
          (Med-Gen) Fact extraction and verification, where models are trained to determine the veracity of claims based on provided evidence. &
          Truthfulness \\
          \hline
            \bottomrule
        \end{tabular}
        \label{tab:Datasets1}\end{table*}

\begin{table*}[htbp]
        \centering
        \small
        \renewcommand{\arraystretch}{1.2}
        \setlength{\tabcolsep}{6pt}
        \begin{tabular}{p{2cm} p{2.5cm} p{4cm} p{3.5cm} p{1.5cm}}
            \toprule
            \textbf{Datasets} & \textbf{Data Type} & \textbf{Content} & \textbf{Task} & \textbf{Dimensions} \\
            \midrule
        \href{https://aclanthology.org/2021.findings-emnlp.297/}{HEALTHVER} &
          Curated Domain-Specific Dataset. &
          contains 14,330 evidence-claim pairs labeled as SUPPORTS, REFUTES, or NEUTRAL, derived from real-world health claims, mainly about COVID-19, verified against scientific articles. &
          (Med-Gen) Training and evaluating models on the task of verifying the truthfulness of health-related claims by assessing their alignment with scientific evidence. This involves classifying claims as supported, refuted, or neutral based on the provided evidence. &
          Truthfulness \\
          \hline
            \href{https://medhalt.github.io/}{Med-HALT} &
          Synthetic and Real Data, Curated Domain-Specific Dataset, and Public Dataset &
          59,254 entries. Consist of Reasoning-Based Assessments, Memory-Based Assessments, Medical Scenarios, Evaluation Metrics &
          (Med-Gen) Tasks including Evaluation of Hallucination in Medical AI, Reliability Benchmarking, Error Analysis, Mitigation Development &
          Truthfulness \\
          \hline
            \href{https://github.com/allenai/medicat}{MedICaT} &
          Public Text Corpora And Real Data (curated from publicly available biomedical literature) &
          217,060 figures extracted from 131,410 open-access papers. Contains medical images (e.g., radiographs, charts, and diagrams) paired with captions extracted from biomedical literature. Also, includes metadata about the source and context of the images. &
          (Med-Gen) Task including Medical Image Captioning, Text-Image Retrieval, Medical Reasoning &
          Truthfulness \\
          \hline
            \href{https://bmcbioinformatics.biomedcentral.com/articles/10.1186/s12859-015-0564-6}{BioASQ} &
          Curated Domain-Specific Dataset; Real Data. &
          3,743 training questions and 500 test questions. The dataset comprises English-language biomedical questions, each accompanied by reference answers and related materials. These questions are designed to reflect real information needs of biomedical experts, making the dataset both realistic and challenging. &
          (Med-QA) The primary task is Biomedical Question Answering (QA), which involves systems providing accurate answers to questions based on biomedical data. The dataset supports various QA tasks, including yes/no, factoid, list, and summary questions. &
          Truthfulness \\
          \hline
            \href{https://pmc.ncbi.nlm.nih.gov/articles/PMC11186750/}{FactualBio} &
          Synthetic Data; Public Text Corpora. &
          collection of biographies of individuals notable enough to have Wikipedia pages but lacking extensive detailed coverage. The dataset was generated using GPT-4 and includes biographies of 21 individuals randomly sampled from the WikiBio dataset. &
          (Med-Gen) Evaluating the factual accuracy of language models, particularly in the context of biography generation. It serves as a benchmark for detecting hallucinations and assessing the factual consistency of generated text. &
          Truthfulness \\
          \hline
            \href{https://pubmedqa.github.io/}{PubMedQA} &
          Curated Domain-Specific Dataset. &
          Consists of over 1,000 question-answer pairs derived from PubMed abstracts, focusing on various biomedical topics. &
          (Med-QA) Evaluates the ability of models to comprehend and extract information from biomedical texts to answer specific questions. &
          Truthfulness \\
          \hline
            \href{https://github.com/abachaa/MedQuAD}{MedQuAD} &
          Curated Domain-Specific Dataset. &
          The dataset encompasses 37 question types, such as Treatment, Diagnosis, and Side Effects, associated with diseases, drugs, and other medical entities like tests. &
          (Med-QA) Designed for medical question answering, the dataset aids in developing and evaluating systems that can understand and respond to medical inquiries. &
          Truthfulness \\
          \hline
            \href{https://github.com/abachaa/LiveQA_MedicalTask_TREC2017}{LiveMedQA2017} &
          Curated Domain-Specific Dataset &
          Consists of 634 question-answer pairs corresponding to National Library of Medicine (NLM) questions &
          (Med-QA) Medical question answering, focusing on consumer health questions received by the U.S. National Library of Medicine. &
          Truthfulness \\
          \hline
            \bottomrule
        \end{tabular}
        \label{tab:Datasets2}\end{table*}

\begin{table*}[htbp]
        \centering
        \small
        \renewcommand{\arraystretch}{1.2}
        \setlength{\tabcolsep}{6pt}
        \begin{tabular}{p{2cm} p{2.5cm} p{4cm} p{3.5cm} p{1.5cm}}
            \toprule
            \textbf{Datasets} & \textbf{Data Type} & \textbf{Content} & \textbf{Task} & \textbf{Dimensions} \\
            \midrule
        \href{https://aclanthology.org/2020.findings-emnlp.342/}{MASH-QA} &
          Curated Domain-Specific Dataset. &
          Approximately 25,000 question-answer pairs sourced from WebMD, covering a wide range of healthcare topics. &
          (Med-QA) Designed for multiple-answer span extraction in healthcare question answering. &
          Truthfulness \\
          \hline
            \href{https://aclanthology.org/2024.findings-emnlp.346.pdf}{SecureSQL} &
          Curated domain-specific dataset &
          Comprises meticulously annotated samples, including both positive and negative instances. The dataset encompasses 57 databases across 34 diverse domains, each associated with specific security conditions. &
          (Med-IE) Evaluate and analyze data leakage risks in LLMs, particularly concerning SQL query generation and execution. &
          Privacy \\
          \hline
            \href{https://arxiv.org/abs/2304.08247}{Medical Meadow} &
          curated domain-specific dataset &
          It comprises approximately 1.5 million data points across various tasks, including question-answer pairs generated from openly available medical data using models like OpenAI's &
          (Med-Gen) Designed to enhance large language models (LLMs) for medical applications &
          Privacy \\
          \hline
            \href{https://www.nature.com/articles/s41598-025-86890-3}{Electronic Health Records (EHR) at (KHCC)} &
          Private dataset &
          gpt-3.5-turbo &
          (Med-IE) Clinical research, outcome analysis. &
          Privacy \\
          \hline
            \href{https://huggingface.co/datasets/agupte/MedVQA}{MedVQA} &
          Curated domain-specific dataset &
          794 image-question-answer triplets. A collection of medical visual question answering pairs, designed to train and evaluate models that interpret medical images and answer related questions. &
          (Med-QA) Visual question answering, medical image understanding. &
          Explainability \\
          \hline
            \href{https://aclanthology.org/2024.bionlp-1.14/}{MedExQA} &
          Curated domain-specific dataset &
          965 multiple-choice medical questions. A dataset focused on medical examination questions and answers, intended to aid in the development of AI models for medical exam preparation and assessment. &
          (Med-QA) Question answering, educational assessment. &
          Explainability \\
          \hline
            \href{https://medmcqa.github.io}{MedMCQA} &
          Curated domain-specific dataset &
          194,000 multiple-choice questions from AIIMS and NEET PG entrance exams, covering 2,400 healthcare topics across 21 medical subjects. A multiple-choice question-answering dataset in the medical domain, aimed at training models to handle medical examinations and practice questions. &
          (Med-QA) Multiple-choice question answering, medical education. &
          Explainability \\
          \hline
            \href{https://arxiv.org/abs/2502.09156}{TCM Medical Licensing Examination(MLE)} &
          Curated domain-specific dataset &
          600 multiple-choice questions. A dataset comprising questions and answers from Traditional Chinese Medicine licensing examinations. &
          (Med-QA) Educational assessment, question answering. &
          Explainability \\
          \hline
            \bottomrule
        \end{tabular}
        \label{tab:Datasets3}\end{table*}

\begin{table*}[htbp]
        \centering
        \small
        \renewcommand{\arraystretch}{1.2}
        \setlength{\tabcolsep}{6pt}
        \begin{tabular}{p{2cm} p{2.5cm} p{4cm} p{3.5cm} p{1.5cm}}
            \toprule
            \textbf{Datasets} & \textbf{Data Type} & \textbf{Content} & \textbf{Task} & \textbf{Dimensions} \\
            \midrule
        \href{https://www.kaggle.com/datasets/paultimothymooney/chest-xray-pneumonia}{Pneumonia Dataset} &
          Curated domain-specific dataset &
          5,863 images. Medical images (such as chest X-rays) labeled for the presence or absence of pneumonia, used for training diagnostic models. &
          (Med-IE) Image classification, disease detection. &
          Explainability \\
          \hline
            \href{https://www.kaggle.com/datasets/raddar/tuberculosis-chest-xrays-montgomery}{Montgomery Dataset} &
          Curated domain-specific dataset &
          X-ray Set comprises 138 posterior-anterior chest X-ray images, with 80 normal and 58 abnormal cases indicative of tuberculosis. Chest X-ray images with manual segmentations of the lung fields, useful for pulmonary research. &
          (Med-IE) Image segmentation, tuberculosis detection. &
          Explainability \\
          \hline
            \href{https://www.kaggle.com/datasets/raddar/tuberculosis-chest-xrays-shenzhen}{Shenzhen Dataset} &
          Curated domain-specific dataset &
          Chest X-ray dataset comprises 662 frontal chest X-rays, including 326 normal cases and 336 cases with manifestations of tuberculosis. Chest X-ray images collected in Shenzhen, China, with annotations for tuberculosis manifestations. &
          (Med-IE) Disease classification, image analysis. &
          Explainability \\
          \hline
            \href{https://ieee-dataport.org/open-access/indian-diabetic-retinopathy-image-dataset-idrid}{IDRID Dataset} &
          Curated domain-specific dataset &
          1,113 images. Retinal images with annotations for diabetic retinopathy lesions, intended for retinal image analysis. &
          (Med-IE) Image segmentation, disease grading. &
          Explainability \\
          \hline
            \href{https://www.nature.com/articles/s41597-022-01899-x}{MIMIC IV} &
          Curated Real-World Clinical Dataset &
          Over 300,000 hospital admissions from Beth Israel Deaconess Medical Center covering de-identified EHR data including demographics, vital signs, medications, diagnoses, and clinical notes &
          (Med-IE / Med-QA / Med-Gen) Used for tasks such as medical code prediction, patient outcome forecasting, clinical summarization, and question answering &
          Explainability \\
          \hline
            \bottomrule
        \end{tabular}
                \caption{This table provides a structured comparison of datasets used in studies on trust in LLMs for healthcare. The datasets are categorized by data type (e.g., web-scraped, curated domain-specific, synthetic, real-world, or private datasets), content (e.g., medical literature, patient records, clinical guidelines, QA pairs), task (e.g., clinical decision support, medical question-answering, document summarization, biomedical fact-checking, chatbot training), and dimensions of trustworthiness (e.g., truthfulness, privacy, safety, robustness, fairness, bias, explainability). This comparison highlights how each dataset contributes to the development of trustworthy LLMs in medical AI.}
\label{tab:Datasets4}\end{table*}

\clearpage
\section{Comparison of Models}
\label{app:gpts}
We systematically gathered and analyzed 81 models relevant to studies on trust in LLMs for healthcare. Table 2 provides a comprehensive summary of the LLMs evaluated in these studies, detailing key aspects such as model name, release year, openness, architecture, and the institution responsible for its development. Additionally, it specifies the primary task each model is designed for, including medical question-answering, clinical decision support, and biomedical text summarization. To further assess their reliability, we categorize the models based on the dimensions of trustworthiness they address, such as truthfulness, privacy, safety, robustness, fairness and bias, and explainability. This structured overview offers valuable insights into how different LLMs are designed and evaluated to enhance trust in healthcare AI applications.

\begin{table*}[htbp]
        \centering
        \small
        \renewcommand{\arraystretch}{1.2}
        \setlength{\tabcolsep}{6pt}
        \begin{tabular}{p{1.8cm} p{1.2cm} p{1.8cm} p{1.6cm} p{1.6cm} p{4.6cm} p{1.9cm}}
            \toprule
            \textbf{Models} & \textbf{Release Year} & \textbf{Institution} & \textbf{Openness} & \textbf{Architecture} & \textbf{Primary Task} & \textbf{Dimensions} \\
            \midrule
        \href{https://github.com/allenai/scibert}{SciBERT} &
          2019 &
          Allen Institute for AI &
          Open-source &
          Encoder-only &
          Pre-trained language model specialized for scientific text, particularly biomedical and computer science literature. &
          Fairness and Bias \\
          \hline
            \href{https://github.com/conceptofmind/PaLM}{PaLM-2} &
          2023 &
          Google &
          Closed-source &
          Decoder-only &
          Multilingual language understanding and generation, with a focus on reasoning and coding tasks. &
          Fairness and Bias \\
          \hline
            \href{https://huggingface.co/mistralai/Mixtral-8x7B-v0.1}{Mixtral-8x70B} &
          2023 &
          Mistral AI &
          Open-source &
          Decoder-only &
          Ensemble of language models aimed at improving performance across diverse language tasks. &
          Fairness and Bias, Safety \\
          \hline
            \href{https://github.com/kyegomez/Med-PaLM}{Med-PaLM} &
          2023 &
          Google Health &
          Closed-source &
          Decoder-only &
          Specializing in healthcare-related question answering, clinical diagnosis support, and medical literature interpretation. &
          Fairness and Bias \\
          \hline
            \href{https://github.com/kyegomez/Med-PaLM}{Med-PaLM 2} &
          2024 &
          Google Health &
          Closed-source &
          Encoder-decoder &
          Updated version of Med-PaLM, further improving healthcare-related tasks with enhanced accuracy and reliability in medical information retrieval, clinical reasoning, and decision support. &
          Fairness and Bias \\
          \hline
            \href{https://github.com/meta-llama/llama}{Llama-13B} &
          2023 &
          Meta &
          Open-source &
          Decoder-only &
          Designed for natural language understanding and generation tasks, such as text summarization, machine translation, and conversational AI. &
          Fairness and Bias \\
          \hline
            \href{https://github.com/zihangdai/xlnet}{XLNet} &
          2019 &
          Google Research &
          Open-source &
          Encoder-only &
          It is used for text classification, question answering, and language modeling tasks. &
          Fairness and Bias \\
          \hline
            \href{https://github.com/microsoft/DeBERTa}{DeBERTa} &
          2020 &
          Microsoft Research &
          Open-source &
          Encoder-only &
          Improves BERT and RoBERTa by enhancing the attention mechanism. It performs well in a variety of NLP tasks, such as sentence classification, question answering, and named entity recognition. &
          Fairness and Bias \\
          \hline
            \href{https://github.com/meta-llama/llama}{Llama-7B} &
          2023 &
          Meta &
          Open-source &
          Decoder-only &
          Focused on general-purpose natural language understanding and generation, with potential fine-tuning for specific domains like medicine, law, and technology. &
          Fairness and Bias, Truthfulness \\
          \hline
            \href{https://huggingface.co/meta-llama/Llama-2-70b-chat-hf}{Llama 2 70Bchat} &
          2023 &
          Meta Platforms &
          Open-source &
          Decoder-only &
          Open-source conversational AI model designed for dialogue and instruction-following tasks. &
          Fairness and Bias, Truthfulness, Safety, Robustness, \\
          \hline
            \href{https://github.com/kydycode/chatgpt-3.5-turbo}{GPT-3.5} &
          2022 &
          OpenAI &
          Closed-source &
          Decoder-only &
          Enhanced language processing capabilities, building upon GPT-3. &
          Fairness and Bias, Truthfulness, Safety, Robustness, Privacy \\
          \hline
            \href{https://github.com/openai/gpt-2}{GPT2} &
          2019 &
          OpenAI &
          Open-source &
          Decoder-only &
          Text generation &
          Fairness and Bias, Robustness \\
          \hline
            \bottomrule
        \end{tabular}
        \label{tab:Models}\end{table*}

\begin{table*}[htbp]
        \centering
        \small
        \renewcommand{\arraystretch}{1.2}
        \setlength{\tabcolsep}{6pt}
        \begin{tabular}{p{1.8cm} p{1.2cm} p{1.8cm} p{1.6cm} p{1.6cm} p{4.6cm} p{1.9cm}}
            \toprule
            \textbf{Models} & \textbf{Release Year} & \textbf{Institution} & \textbf{Openness} & \textbf{Architecture} & \textbf{Primary Task} & \textbf{Dimensions} \\
            \midrule
        \href{https://github.com/chaoyi-wu/PMC-LLaMA}{PMC Llama 13B} &
          2023 &
          Allen Institute for AI &
          Open-source &
          Decoder-only &
          Specialized in medical literature understanding and generation. &
          Fairness and Bias, Robustness \\
          \hline
            \href{https://openai.com/index/gpt-4/}{GPT-4} &
          2023 &
          OpenAI &
          Closed-source &
          Decoder-only &
          Advanced language generation and understanding across various domains. &
          Fairness and Bias, Safety, Robustness, Explainability, Privacy \\
          \hline
            \href{https://huggingface.co/docs/transformers/en/model_doc/bert}{BERT} &
          2018 &
          Google AI Language &
          Open-source &
          Encoder-only &
          Pre-trained Transformer model for a wide range of NLP tasks, such as text classification, NER, QA, etc. &
          Fairness and Bias, Safety, Robustness, Truthfulness \\
          \hline
            \href{https://github.com/dataprofessor/llama2}{LLAMA 2 CHAT} &
          2023 &
          Meta AI &
          Open-source &
          Decoder-only &
          Language modeling &
          Robustness, Explainability \\
          \hline
            \href{https://huggingface.co/medalpaca/medalpaca-7b}{MEDALPACA (7B)} &
          2023 &
          medalpaca &
          Open-source &
          Decoder-only &
          Medical domain language model fine-tuned for question-answering and medical dialogue tasks. &
          Robustness, Privacy \\
          \hline
            \href{https://github.com/bowang-lab/clinical-camel}{CLINICAL CAMEL (13B)} &
          2023 &
          the AI and healthcare community &
          Open-source &
          Decoder-only &
          Fine-tuned for clinical applications. It is designed to assist with tasks like medical text classification, clinical decision support, information extraction from medical records, and answering clinical questions. &
          Robustness \\
          \hline
            \href{https://huggingface.co/openai-community/gpt2-xl}{GPT-2 XL} &
          2019 &
          OpenAI &
          Open-source &
          Decoder-only &
          Large-scale language model for text generation and understanding. &
          Robustness \\
          \hline
            \href{https://huggingface.co/google-t5/t5-large}{T5-Large} &
          2020 &
          Google Research &
          Open-source &
          Encoder-decoder &
          It treats all NLP tasks as text-to-text tasks, meaning both the input and output are in the form of text, and it's used for tasks like translation, summarization, and question answering. &
          Robustness \\
          \hline
            \href{https://www.anthropic.com/news/claude-3-5-sonnet}{claude-3.5-sonnet} &
          2024 &
          Anthropic &
          Closed-source &
          Decoder-only &
          It is a variant of Claude, specialized in tasks such as conversational AI, creative writing, poetry generation, and other text-based applications. &
          Robustness \\
          \hline
            \href{https://huggingface.co/aaditya/Llama3-OpenBioLLM-70B}{OpenBioLLM-70B} &
          2024 &
          OpenBioAI &
          Open-source &
          Decoder-only &
          It is designed to handle tasks such as biological information extraction, gene sequence analysis, protein folding predictions, and other bioinformatics applications. &
          Robustness \\
          \hline
            \href{https://huggingface.co/BioMistral/BioMistral-7B}{BioMistral-7B} &
          2023 &
          Mistral AI &
          Open-source &
          Decoder-only &
          Focused on biomedical and healthcare-related text. Its tasks include medical question answering, clinical document analysis, and medical text summarization. &
          Robustness \\
          \hline
            \href{https://huggingface.co/ProbeMedicalYonseiMAILab/medllama3-v20}{Medllama3-v20} &
          2024 &
          MedAI Labs &
          Open-source &
          Decoder-only &
          Designed to assist in healthcare tasks like clinical reasoning, medical question answering, and patient record analysis. &
          Robustness \\
          \hline
            \bottomrule
        \end{tabular}
        \label{tab:Models1}\end{table*}

\begin{table*}[htbp]
        \centering
        \small
        \renewcommand{\arraystretch}{1.2}
        \setlength{\tabcolsep}{6pt}
        \begin{tabular}{p{1.8cm} p{1.2cm} p{1.8cm} p{1.6cm} p{1.6cm} p{4.6cm} p{1.9cm}}
            \toprule
            \textbf{Models} & \textbf{Release Year} & \textbf{Institution} & \textbf{Openness} & \textbf{Architecture} & \textbf{Primary Task} & \textbf{Dimensions} \\
            \midrule
        \href{https://huggingface.co/starmpcc/Asclepius-Llama2-7B}{ASCLEPIUS (7B)} &
          2023 &
          Asclepius AI &
          Open-source &
          Decoder-only &
          Developed for clinical and medical applications, specializing in tasks like diagnosing medical conditions from symptoms, medical text summarization, and extracting structured information from clinical documents. &
          Robustness, Explainability \\
          \hline
            \href{https://github.com/tatsu-lab/stanford_alpaca}{ALPACA (7B)} &
          2023 &
          Stanford University &
          Open-source &
          Decoder-only &
          Fine-tuned version of the LLaMA model aimed at providing high-quality responses to questions, with an emphasis on maintaining ethical and accurate conversational capabilities in diverse domains. &
          Robustness \\
          \hline
            \href{https://github.com/ra83205/google-bard-api}{Google’s Bard} &
          2023 &
          Google &
          Closed-source &
          Encoder-decoder &
          Conversational AI tool, focused on providing detailed, accurate, and creative responses to user queries. It can handle a variety of tasks, including web search, content generation, and complex QA. &
          Robustness \\
          \hline
            \href{https://github.com/gabrielsants/openai-davinci-003}{Text- Davinci-003} &
          2022 &
          OpenAI &
          Closed-source &
          Decoder-only &
          It is an advanced variant of GPT-3. It is designed for a wide range of natural language understanding and generation tasks, such as answering questions, summarizing text, creative writing, translation, and code generation. &
          Robustness, Truthfulness \\
          \hline
            \href{https://huggingface.co/meta-llama/Llama-2-7b}{LLaMa 2-7B} &
          2023 &
          Meta (formerly Facebook AI Research) &
          Open-source &
          Decoder-only &
          Designed to be a general-purpose AI for a wide range of tasks such as text generation, question answering, and summarization, with specific fine-tuning for medical and technical domains. &
          Robustness, Truthfulness, Privacy \\
          \hline
            \href{https://openai.com/index/chatgpt/}{ChatGPT} &
          2022 &
          OpenAI &
          Closed-source &
          Decoder-only &
          Conversational AI &
          Robustness, Truthfulness, Explainability, Privacy \\
          \hline
            \href{https://github.com/meta-llama/llama3}{Llama-3.1} &
          2024 &
          Meta AI &
          Open-source &
          Decoder-only &
          Multilingual large language model designed for a variety of natural language processing tasks. &
          Safety, privacy \\
          \hline
            \href{https://huggingface.co/wanglab/ClinicalCamel-70B}{ClinicalCamel-70b} &
          2023 &
          the AI and healthcare community &
          Open-source &
          Decoder-only &
          Medical language model designed for clinical research applications. &
          Safety, Explainability \\
          \hline
            \href{https://github.com/m42-health/med42}{Med42-70b} &
          2023 &
          M42 Health &
          Open-source &
          Decoder-only &
          Clinical large language model providing high-quality answers to medical questions. &
          Safety, Explainability \\
          \hline
            \href{https://openai.com/index/hello-gpt-4o/}{GPT-4o} &
          2024 &
          OpenAI &
          Closed-source &
          Decoder-only &
          Multimodal large language model capable of processing and generating text, audio, and images in real time. &
          Safety, Privacy, Explainability \\
          \hline
            \href{https://github.com/mistralai/mistral-inference}{Mistral} &
          2023 &
          Mistral AI &
          Open-source &
          Decoder-only &
          Language model optimized for code generation and reasoning tasks. &
          Safety, Robustness, Explainability \\
          \hline
            \href{https://github.com/epfLLM/meditron}{Meditron (7) (70b)} &
          2023 &
          École Polytechnique Fédérale de Lausanne (EPFL) &
          Open-source &
          Decoder-only &
          Medical language model fine-tuned for clinical decision support and medical reasoning. &
          Safety, Robustness, Explainability \\
          \hline
            \bottomrule
        \end{tabular}
        \label{tab:Models2}\end{table*}

\begin{table*}[htbp]
        \centering
        \small
        \renewcommand{\arraystretch}{1.2}
        \setlength{\tabcolsep}{6pt}
        \begin{tabular}{p{1.8cm} p{1.2cm} p{1.8cm} p{1.6cm} p{1.6cm} p{4.6cm} p{1.9cm}}
            \toprule
            \textbf{Models} & \textbf{Release Year} & \textbf{Institution} & \textbf{Openness} & \textbf{Architecture} & \textbf{Primary Task} & \textbf{Dimensions} \\
            \midrule
        \href{https://www.anthropic.com/news/claude-2-1}{Claude-2.1} &
          2023 &
          Anthropic &
          Closed-source &
          Decoder-only &
          General-purpose language model for a wide range of natural language understanding and generation tasks. &
          Safety, Robustness \\
          \hline
            \href{https://huggingface.co/docs/transformers/en/model_doc/gptj}{GPT-J} &
          2021 &
          EleutherAI &
          Open-source &
          Decoder-only &
          Open-source language model for text generation and understanding. &
          Safety, Robustness \\
          \hline
            \href{https://github.com/eddieali/Vicuna-AI-LLM}{Vicuna} &
          2023 &
          UC Berkeley and Microsoft Research &
          Open-source &
          Decoder-only &
          Conversational AI &
          Safety, Robustness, Truthfulness \\
          \hline
            \href{https://huggingface.co/medalpaca/medalpaca-13b}{Medalpaca-13b} &
          2023 &
          medalpaca &
          Open-source &
          Decoder-only &
          Medical domain language model fine-tuned for question-answering and medical dialogue tasks. &
          Safety, Truthfulness, Privacy \\
          \hline
            \href{https://openai.com/index/gpt-3-apps/}{GPT-3} &
          2020 &
          OpenAI &
          Closed-source &
          Decoder-only &
          Natural language understanding and generation &
          Truthfulness, Explainability \\
          \hline
            \href{https://github.com/google-research/albert}{ALBERT} &
          2019 &
          Google Research &
          Open-source &
          Encoder-only &
          Lighter version of BERT that reduces parameters for efficiency while maintaining performance. It excels in tasks such as text classification, named entity recognition, and question answering. &
          Truthfulness \\
          \hline
            \href{https://huggingface.co/docs/transformers/en/model_doc/roberta}{RoBERTa} &
          2019 &
          Facebook AI Research &
          Open-source &
          Encoder-only &
          Optimized variant of BERT that removes the Next Sentence Prediction task and trains with more data and for longer periods. It is used for tasks like question answering, sentiment analysis, and text classification. &
          Truthfulness \\
          \hline
            \href{https://github.com/ncbi-nlp/bluebert}{BlueBERT} &
          2019 &
          NIH and Stanford University &
          Open-source &
          Encoder-only &
          BERT-based model pre-trained on clinical and biomedical text. It is designed for healthcare-related tasks, including clinical text classification, named entity recognition, and medical question answering. &
          Truthfulness \\
          \hline
            \href{https://github.com/kexinhuang12345/clinicalBERT}{ClinicalBERT} &
          2019 &
          University of Pennsylvania &
          Open-source &
          Encoder-only &
          Variant of BERT fine-tuned on clinical texts, tailored for clinical NLP tasks like named entity recognition, clinical event extraction, and question answering in the medical domain. &
          Truthfulness \\
          \hline
            \href{https://github.com/google-research/tapas}{TAPAS} &
          2020 &
          Google Research &
          Open-source &
          Encoder-only &
          Designed for answering questions based on tabular data. It is used for tasks like extracting structured information from tables and processing queries in tabular datasets. &
          Truthfulness \\
          \hline
            \href{https://huggingface.co/meta-llama/Llama-2-13b}{LLaMA-2 13B} &
          2023 &
          Meta &
          Open-source &
          Decoder-only &
          Advanced variant of Meta's LLaMA series, designed for text generation, question answering, summarization, and other NLP tasks. &
          Truthfulness, Explainability, Privacy \\
          \hline
            \href{https://github.com/mosaicml/llm-foundry}{MPT} &
          2023 &
          MosaicML &
          Open-source &
          Decoder-only &
          General-purpose LLM for text generation, summarization, language understanding, and reasoning tasks. Fine-tuned for downstream applications such as chatbot development, code generation, and other NLP tasks. &
          Truthfulness \\
          \hline
            \bottomrule
        \end{tabular}
        \label{tab:Models3}\end{table*}

\begin{table*}[htbp]
        \centering
        \small
        \renewcommand{\arraystretch}{1.2}
        \setlength{\tabcolsep}{6pt}
        \begin{tabular}{p{1.8cm} p{1.2cm} p{1.8cm} p{1.6cm} p{1.6cm} p{4.6cm} p{1.9cm}}
            \toprule
            \textbf{Models} & \textbf{Release Year} & \textbf{Institution} & \textbf{Openness} & \textbf{Architecture} & \textbf{Primary Task} & \textbf{Dimensions} \\
            \midrule
        \href{https://huggingface.co/docs/transformers/main/model_doc/blip-2}{BLIP2} &
          2023 &
          Salesforce &
          Open-source &
          Encoder-decoder &
          Bootstrapping language-image pre-training, designed to bridge vision-language models with large language models for improved visual understanding and generation. &
          Truthfulness \\
          \hline
            \href{https://huggingface.co/Salesforce/instructblip-vicuna-7b?utm_source=chatgpt.com}{InstructBLIP-7b/13b} &
          2023 &
          Salesforce &
          Open-source &
          Encoder-decoder &
          Visual instruction-tuned versions of BLIP-2, utilizing Vicuna-7B and Vicuna-13B language models, respectively, to enhance vision-language understanding through instruction tuning. &
          Truthfulness \\
          \hline
            \href{https://github.com/haotian-liu/LLaVA}{LLaVA1.5-7b/13b} &
          2023 &
          Microsoft &
          Open-source &
          Encoder-decoder &
          Large language and vision assistant models with 7B and 13B parameters, respectively, designed for multimodal tasks by integrating visual information into language models. &
          Truthfulness \\
          \hline
            \href{https://github.com/X-PLUG/mPLUG-Owl?tab=readme-ov-file}{mPLUGOwl2} &
          2023 &
          Zhejiang University &
          Open-source &
          Encoder-decoder &
          Multimodal pre-trained language model designed to handle various vision-language tasks, including image captioning and visual question answering. &
          Truthfulness \\
          \hline
            \href{https://github.com/mbzuai-oryx/XrayGPT}{XrayGPT} &
          2023 &
          University of Toronto &
          Open-source &
          Decoder-only &
          Specialized model for generating radiology reports from chest X-ray images, aiming to assist in medical image interpretation. &
          Truthfulness \\
          \hline
            \href{https://github.com/Vision-CAIR/MiniGPT-4}{MiniGPT4} &
          2023 &
          King Abdullah University of Science and Technology &
          Open-source &
          Decoder-only &
          A lightweight multimodal model designed to align vision and language models efficiently, facilitating tasks like image captioning and visual question answering. &
          Truthfulness \\
          \hline
            \href{https://github.com/chaoyi-wu/RadFM}{RadFM} &
          2023 &
          Stanford University &
          Open-source &
          Decoder-only &
          Foundation model tailored for radiology, focusing on interpreting medical images and integrating findings with clinical language models. &
          Truthfulness \\
          \hline
            \href{https://github.com/tloen/alpaca-lora}{Alpaca-LoRA} &
          2023 &
          Stanford University &
          Open-source &
          Decoder-only &
          It focuses on achieving good performance in tasks such as question answering and personalized dialogue. &
          Truthfulness \\
          \hline
            \href{https://github.com/Integral-Healthcare/robin-ai-reviewer}{Robin- medical} &
          2023 &
          Robin Health &
          Open-source &
          Decoder-only &
          Fine-tuned for medical applications, including clinical decision support, medical question answering, and health record analysis. &
          Truthfulness \\
          \hline
            \href{https://huggingface.co/docs/transformers/en/model_doc/flan-t5}{Flan-T5} &
          2021 &
          Google Research &
          Open-source &
          Encoder-decoder &
          Optimized for tasks like question answering, text summarization, and sentence classification, across a variety of domains. &
          Truthfulness, Explainability \\
          \hline
            \href{https://github.com/dmis-lab/biobert?tab=readme-ov-file}{BioBERT} &
          2019 &
          Korea University &
          Open-source &
          Encoder-only &
          Biomedical language representation learning, enhancing performance on tasks like named entity recognition, relation extraction, and question answering within the biomedical domain. &
          Truthfulness \\
          \hline
            \href{https://github.com/falconry/falcon}{Falcon Instruct (7B and 40B)} &
          2023 &
          Technology Innovation Institute (TII), UAE. &
          Open-source &
          Decoder-only &
          Instruction-tuned language model designed to follow user instructions effectively. &
          Truthfulness, Robustness \\
          \hline
            \bottomrule
        \end{tabular}
        \label{tab:Models4}\end{table*}

\begin{table*}[htbp]
        \centering
        \small
        \renewcommand{\arraystretch}{1.2}
        \setlength{\tabcolsep}{6pt}
        \begin{tabular}{p{1.8cm} p{1.2cm} p{1.8cm} p{1.6cm} p{1.6cm} p{4.6cm} p{1.9cm}}
            \toprule
            \textbf{Models} & \textbf{Release Year} & \textbf{Institution} & \textbf{Openness} & \textbf{Architecture} & \textbf{Primary Task} & \textbf{Dimensions} \\
            \midrule
        \href{https://huggingface.co/mistralai/Mistral-7B-Instruct-v0.3}{Mistral Instruct (7B)} &
          2023 &
          Mistral AI &
          Open-source &
          Decoder-only &
          Instruction-tuned language model designed to follow user instructions effectively. &
          Truthfulness, Robustness \\
          \hline
            \href{https://github.com/falconry/falcon}{Falcon} &
          2023 &
          Technology Innovation Institute (TII), UAE. &
          Open-source &
          Decoder-only &
          General-purpose language model optimized for text understanding, generation, question answering, and reasoning tasks. Focused on efficient deployment for industry-scale applications. &
          Truthfulness, Robustness \\
          \hline
            \href{https://github.com/microsoft/LLaVA-Med?utm_source=chatgpt.com}{LLaVA-Med} &
          2024 &
          Microsoft &
          Open-source &
          Encoder-decoder &
          Large language and vision assistant for biomedicine, trained to handle visual instruction tasks in the biomedical field, aiming for capabilities similar to GPT-4. &
          Truthfulness, Explainability \\
          \hline
            \href{https://claude.ai/login?returnTo=%2F%3F}{Claude-3} &
          2024 &
          Anthropic &
          Closed-source &
          Decoder-only &
          General-purpose LLM (QA, dialogue, reasoning, summarization) &
          Explainability \\
          \hline
            \href{https://openai.com/index/gpt-4o-mini-advancing-cost-efficient-intelligence/}{GPT-4o-mini} &
          2024 &
          OpenAI &
          Closed-source &
          Decoder-only &
          Natural language processing (NLP), text generation, and understanding. &
          Explainability \\
          \hline
            \href{https://huggingface.co/starmpcc/Asclepius-13B}{ASCLEPIUS (13B)} &
          2023 &
          Asclepius AI &
          Open-source &
          Decoder-only &
          Medical NLP, clinical text analysis, and healthcare-related tasks. &
          Explainability \\
          \hline
            \href{https://github.com/MedHK23/MedViLaM}{MedViLaM} &
          2023 &
          Cite &
          Open-source &
          Encoder-decoder &
          Medical vision-language tasks, combining image and text analysis for healthcare. &
          Explainability \\
          \hline
            \href{https://github.com/jiangsongtao/Med-MoE}{Med-MoE} &
          2023 &
          Cite &
          Open-source &
          Decoder-only &
          Medical NLP, leveraging Mixture of Experts (MoE) for specialized healthcare tasks. &
          Explainability \\
          \hline
            \href{https://deepmind.google/technologies/gemini/pro/}{Gemini Pro} &
          2023 &
          Google DeepMind &
          Closed-source &
          Decoder-only &
          Multi-modal NLP, combining text, image, and other data types for advanced AI tasks &
          Explainability \\
          \hline
            \href{https://gemini.google.com/app}{Gemini-1.5} &
          2024 &
          Google DeepMind &
          Closed-source &
          Decoder-only &
          Multimodal reasoning, long-context understanding, QA, generation &
          Explainability \\
          \hline
            \href{https://github.com/XZhang97666/AlpaCare}{AlpaCare (7B) (13B)} &
          2023 &
          Cite &
          Open-source &
          Decoder-only &
          Healthcare-focused NLP, clinical text analysis, and medical decision support &
          Explainability \\
          \hline
            \href{https://huggingface.co/01-ai/Yi-6B}{Yi (6B)} &
          2023 &
          01.AI (China) &
          Open-source &
          Decoder-only &
          General-purpose NLP, text generation, and fine-tuning for specific applications. &
          Explainability \\
          \hline
            \bottomrule
        \end{tabular}
        \label{tab:Models5}\end{table*}

\begin{table*}[htbp]
        \centering
        \small
        \renewcommand{\arraystretch}{1.2}
        \setlength{\tabcolsep}{6pt}
        \begin{tabular}{p{1.8cm} p{1.2cm} p{1.8cm} p{1.6cm} p{1.6cm} p{4.6cm} p{1.9cm}}
            \toprule
            \textbf{Models} & \textbf{Release Year} & \textbf{Institution} & \textbf{Openness} & \textbf{Architecture} & \textbf{Primary Task} & \textbf{Dimensions} \\
            \midrule
        \href{https://huggingface.co/microsoft/phi-2}{Phi-2 (2.7B)} &
          2023 &
          Microsoft &
          Open-source &
          Decoder-only &
          Lightweight NLP, text generation, and fine-tuning for specific tasks. &
          Explainability \\
          \hline
            \href{https://huggingface.co/upstage/SOLAR-10.7B-v1.0}{SOLAR (10.7B)} &
          2023 &
          Upstage AI &
          Open-source &
          Decoder-only &
          General-purpose NLP, text generation, and fine-tuning for specific domains. &
          Explainability \\
          \hline
            \href{https://github.com/InternLM/InternLM}{InternLM2 (7B)} &
          2023 &
          Shanghai AI Laboratory (China) &
          Open-source &
          Decoder-only &
          General-purpose NLP, text generation, and fine-tuning for specific applications. &
          Explainability \\
          \hline
            \href{https://ai.meta.com/blog/meta-llama-3/}{Llama3-( 8B and 70B)} &
          2024 &
          Meta &
          Open-source &
          Decoder-only &
          General-purpose NLP, text generation, and fine-tuning for specific applications. &
          Privacy, Explainability \\
          \hline
            \href{https://huggingface.co/codellama}{CodeLlama-( 7B, 13B, and 34B)} &
          2023 &
          Meta &
          Open-source &
          Decoder-only &
          Code generation, code completion, and programming assistance. &
          Privacy \\
          \hline
            \href{https://mistral.ai/en/news/mixtral-8x22b}{Mixtral-8x7B and 8x22B} &
          2023 &
          Mistral AI &
          Open-source &
          Decoder-only &
          General-purpose NLP, text generation, and fine-tuning for specific domains. &
          Privacy \\
          \hline
            \href{https://huggingface.co/Qwen/Qwen1.5-72B-Chat}{Qwen-(7B, 14B, 32B, 72B)-Chat} &
          2023 &
          Alibaba &
          Open-source &
          Decoder-only &
          Chat-oriented NLP, conversational AI, and text generation. &
          Privacy \\
          \hline
            \href{https://open.bigmodel.cn/dev/api/normal-model/glm-4}{GLM-4} &
          2024 &
          Tsinghua University &
          Open-source &
          Encoder-decoder &
          Advanced NLP, text generation, and multi-modal tasks. &
          Privacy \\
          \hline
            \bottomrule
        \end{tabular}
                \caption{Detailed Comparison of GPT Models Evaluated for Trust in Healthcare LLMs, Including Model Name, Release Year, Institution, Openness, Architecture, Primary Tasks (e.g., Medical Question-Answering, Clinical Decision Support, Biomedical Text Summarization, Medical Report Generation), and Key Trustworthiness Dimensions (Truthfulness, Privacy, Safety, Robustness, Fairness and Bias, Explainability).}
            \label{tab:Models6}\end{table*}

\end{document}